\documentclass{aa}

\usepackage{color}
\usepackage[normalem]{ulem}
\usepackage{amsmath,amssymb}	
\usepackage{textcomp}
\usepackage{booktabs}
\usepackage{xcolor}
\usepackage{nicefrac}
\usepackage{bm}
\usepackage{graphicx}
\usepackage{url}
\usepackage{savesym}
\usepackage[T1]{fontenc}
\usepackage{adjustbox}
\usepackage{lastpage}

\makeatletter
\renewcommand*\aa@pageof{, page \thepage{} of \pageref*{LastPage}}
\makeatother

\usepackage{natbib,twoopt}
\usepackage[caption=false]{subfig}

\usepackage[breaklinks=true, colorlinks=true, linkcolor=black, citecolor=blue, urlcolor=blue]
{hyperref} 
\bibpunct{(}{)}{;}{a}{}{,}

\definecolor{green1}{RGB}{0, 128, 0}

\begin{document} 

    \title{Multidimensional half-moment multigroup radiative transfer}

  \titlerunning{Multidimensional half-moment multigroup radiative transfer}
 \authorrunning{Melon Fuksman, Flock, Klahr, Mattia, Muley}

  \subtitle{Improving moment-based thermal models of circumstellar disks}

     \author{David Melon Fuksman
          \and
          Mario Flock
          \and
          Hubert Klahr
          \and
          Giancarlo Mattia
          \and
          Dhruv Muley
          }

   \institute{Max Planck Institute for Astronomy, K\"onigstuhl 17, 69117 Heidelberg, Germany\\\email{fuksman@mpia.de}}

   \date{Received 1 April 2025 / Accepted July 23 2025}

\abstract
{
Common moment-based radiative transfer methods, such as flux-limited diffusion (FLD) and the M1 closure, suffer from artificial interactions between crossing beams. In protoplanetary disks, this leads to an overestimation of the midplane temperature due to the merging of vertical inward and outward fluxes. Methods that avoid these artifacts typically require angular discretization, which can be computationally expensive.
}
{
In the spirit of the two-stream approximation, we aim to remove the interaction between beams in a fixed spatial direction by introducing a half-moment (HM) closure, which integrates the radiative intensity over hemispheres.
}
{
We derived a multidimensional HM closure via entropy maximization and replaced this closure with an approximate expression that closely matches it, coinciding in the diffusion and free-streaming regimes while remaining expressible through simple operations.
We implemented the HM and M1 closures via implicit-explicit (IMEX) schemes,  including multiple frequency groups.
We tested these methods in numerical benchmarks, such as computing the temperature in an irradiated disk around a T Tauri star, comparing the results with Monte Carlo (MC) radiative transfer simulations.
}
{
The resulting HM closure tends to the correct limit in the diffusion regime and prevents interactions between crossing fluxes in a chosen spatial direction. Using multiple frequency groups, our new method closely reproduces the midplane temperature distributions obtained with classical MC methods in disk simulations. With the M1 closure and a single frequency group, the midplane temperature is around 44\% higher compared to MC.
With 22 frequency groups, the M1 closure agrees with MC by up to $21\%$, while HM reduces this discrepancy to $6\%$. Even with just three frequency groups, HM significantly outperforms M1, with maximum departures of $8\%$ compared to M1's $23\%$. Our results show that combining HM with a multigroup treatment yields more realistic disk temperatures than M1, particularly in optically thick regions.
}
{}


\maketitle

\section{Introduction}\label{S:Introduction}

\nolinenumbers

Astrophysical models often require radiative transfer to properly account for heating and cooling processes. This is generally challenging, as the radiative transfer equation (RTE) depends not only on space and time but also on the direction and energy of the transported photons. Moreover, it is often necessary to model the radiation and matter distributions simultaneously. In this context, many radiative transfer methods rely on simplified moment-based methods, which involve averaging the radiative intensity over all angles and often over frequency as well. Widely used examples include flux-limited diffusion \citep[FLD;][]{Levermore1981FLD}, which involves a single equation for the total radiation energy density, and the two-moment M1 closure \citep[][]{Levermore1984M1}, which introduces an extra equation for the radiation flux.

While these methods often suffice in the diffusion regime, both the angle- and frequency-averaging strategies have drawbacks that can become significant depending on the problem.
Frequency averaging may lead to inaccuracies when dealing with opacities that vary significantly with frequency \citep[see, e.g.,][]{Dullemond2002}. Furthermore, when angle-averaging the radiation intensity, physically motivated ad hoc assumptions must be made in order to close the resulting system of equations, which also introduces inaccuracies. In particular, FLD 
becomes excessively diffusive in the free-streaming regime, where it is unable to cast shadows \citep{Hayes2003}. While the M1 closure does not suffer from this issue, it remains a hydrodynamical method for photon transport, and thus cannot propagate radiation in multiple directions at the same point in space. Therefore, both FLD and M1 lead to artificial interactions between crossing beams \citep[e.g.,][]{Weih2020}, causing inaccuracies wherever such interactions become physically relevant.

These issues can be avoided by applying more refined methods that better capture the angular dependence of the radiation field. This includes approaches relying on an angular discretization of the radiation specific intensity, such as discrete ordinates methods \citep{Jiang2021,Jiang2022,Ma2025}, including long- \citep{Magic2013,Frostholm2018,Stein2024} and short-characteristics \citep{Davis2012} methods—sometimes integrated within variable Eddington tensor approaches \citep{JiangStone2012,Jiang2014}— lattice Boltzmann methods \citep{Asinari2010,Weih2020LatticeBoltzmann}, and reverse ray tracing \citep{Wuensch2021}, to name a few. Recent reviews of angle-discretized solvers can be found in \cite{Wuensch2024} and \cite{Ma2025}. Due to either the additional angular dimension or the iterative nature of some of these approaches, these are generally more computationally expensive than moment methods \citep{Ma2025}, and their applicability depends on the required spatial, angular, and frequency resolution. Some examples of applications to frequency-averaged radiative transfer in the context of planet-forming disks can be found, for example, in \cite{BaileyStoneFung2024} (circumplanetary disks) and \cite{Zhang2024} (circumstellar disks).

Another alternative is provided by Monte Carlo (MC) methods, which involve the propagation of a large number of photon packets through an interacting medium. These schemes naturally account for the frequency and angular dependence of the radiation field and are highly accurate at computing temperature distributions in stationary cases. They are also ideal tools for reproducing scattering-dependent effects, as well as generating synthetic observations from simulations \citep[e.g.,][]{BarrazaAlfaro2024}. Coupling MC temperature computations with hydrodynamics (HD) or magnetohydrodynamics (MHD) via operator splitting is not trivial, and may introduce unphysical phenomena if the timescales of radiative and hydrodynamical processes are altered \citep{MelonFuksman2022}. Instead, a viable option is to use MC methods to compute moments of the radiative intensity, which can then be naturally coupled with the HD/MHD equations \citep{Foucart2018,Smith2020}.

Generally, the high cost of MC methods compared to the previously described approaches \citep{Ma2025}, especially in problems of high optical depth, makes them typically best suited for post-processing computations rather than dynamical simulations \citep[see also][]{Wuensch2024}. Yet, unlike dynamical simulations, such computations do not account for heating and cooling terms that depend on the velocity field, such as expansion and compression work. These processes cannot simply be included as heat sources in MC simulations, as this would leave out local cooling processes. As a result, even in stationary systems, MC post-processing temperature computations can be inaccurate near sources of significant compression and expansion, such as strong spiral waves in circumstellar disks \citep{Krieger2025}.

In standard benchmarks of starlight-irradiated circumstellar disks, several studies show minimal differences between the temperatures obtained with moment-based and MC methods \citep{Pascucci2004,Flock2013,MelonFuksman2021}. However, in largely optically thick disk regions, \cite{Krieger2025} obtained larger midplane temperatures with FLD than with MC \citep[see also][]{MignonRisse2020}. Besides the frequency averaging in FLD, a major reason behind this discrepancy is the fact that the flux entering the disk from the starlight-heated upper layers crosses the flux leaving the disk and cooling it down \citep{MelonFuksman2022}. In moment methods, this results in an artificial interaction between the outward and inward fluxes, reducing the disk's cooling efficiency. This inspires the goal of this paper: to prevent this interaction and develop a method suitable for dynamical simulations that produces accurate disk temperatures with minimal angular and frequency discretization.

To remove the interaction between photons entering and leaving the disk, it is immediate to think of the two-stream approximation, originally developed for stellar atmospheres \citep{Schuster1905,Eddington1916} and widely used in 1D planetary atmosphere models \citep{Malik2017,Mukherjee2023}. This method replaces the radiative intensity with two fields corresponding to its integral in opposite halves of a plane; in other words, separate ``upward'' and ``downward'' fluxes are considered. In the context of circumstellar disks, we can expect that using two different fields for radiation entering and leaving the disk should prevent the artificial interaction produced in moment-based methods and yield more accurate temperature distributions.

A class of two-stream methods is given by half-moment (HM) closures, first introduced for 1D transport in \cite{Dubroca2002} and extended to multigroup radiative transfer in \cite{Turpault2004}. These methods are similar to moment-based closures, with the difference that the solid-angle integration of the radiation specific intensity is replaced with two integrals over hemispheres (half moments). The resulting system of equations is closed by imposing a maximum entropy constraint on the resulting fields, just as the M1 closure can be obtained using a closed angle integration. As in the M1 case, the obtained closure can be expressed in terms of simple relations between the involved half moments. The resulting transport equations are hyperbolic and can be solved using appropriate standard methods.

A generalization of HM methods to multidimensional transport is not straightforward, and few attempts have been made. The case of 2D transport was approached in \cite{Frank2006} by using partial moments integrated over four quadrants. The resulting closure requires a numerical inversion between Lagrange multipliers and partial moments, leading to obvious limitations in computational efficiency. Alternatively, \cite{Ripoll2005} derived a 3D HM closure by assuming that (I) the partial fluxes are always parallel to the total flux and (II) the radiation specific intensity is isotropic in the plane perpendicular to the total flux. Though suitable for the diffusion regime, these approximations fail in cases involving obliquely crossing beams.

In this work, we introduce an HM closure suitable for multidimensional radiative transport. This closure can be derived by replacing the exact calculation of half moments via entropy maximization with an approximate functional form that reproduces the exact maximum-entropy closure with high accuracy, coinciding with it in the free-streaming and diffusion regimes. The half moments are defined by integrating over two halves of a plane whose normal, henceforth the ``splitting direction'', is fixed in time. We also extend this closure to multigroup transport \citep[see, e.g.,][]{Rosdahl2013,Gonzalez2015}; specifically, we divide the frequency domain into a series of discrete bins and solve separate HM equations for each of them, coupled only by radiation-matter interaction. As is shown later, this extension is essential to reproduce vertical temperature variations in irradiated disks.

We implemented these methods as extensions of the M1 radiative transfer code by \cite{MelonFuksman2021}, integrated into the open-source PLUTO code \citep{Mignone2007} for HD and MHD. We employed a finite-volume scheme to compute the transport of the half moments, integrating the resulting system of equations via implicit-explicit (IMEX) methods \citep{PareschiRusso2005}. This approach removes the need for globally implicit solvers, which is generally advantageous for parallelization. The closure is simple enough to compute the eigenvalues of the HM block exactly, which we used to derive and implement Riemann solvers.

For nonrelativistic problems, the light-crossing time of one grid cell is several orders of magnitude smaller than the fluid-crossing time, leading to a Courant-Friedrichs-Lewy (CFL) time step criterion that makes explicit integration prohibitively expensive. One way around this problem, which we employ in the present work, is to replace the speed of light, $c$, with an artificially reduced value, $\hat{c}$ \citep[e.g.,][]{GnedinAbel2001}, which reproduces the correct physical limit provided that $\hat{c}$ is nevertheless significantly larger than all relevant velocities in the problem. Alternatively, one could use a globally implicit approach incorporating both transport and source terms, avoiding the CFL limit.

This article is organized as follows. In Sect. \ref{S:Motivation} we introduce general maximum-entropy closures and define the closure implemented in this work. In Sect. \ref{S:HalfMoment} we summarize the systems of equations considered for HM and M1 radiative transfer. In Sect. \ref{S:NumericalImpl} we describe the numerical implementation of our methods, and in Sect. \ref{S:Tests} we test the code’s performance on a series of numerical benchmarks. In Sect. \ref{S:Disks}, we test the performance of the HM and M1 methods in modeling the temperature of passive irradiated protoplanetary disks in comparison with MC simulations. In Sect. \ref{S:Discussion} we discuss the accuracy and caveats of these disk models and characterize our methods' computational efficiency. Finally, in Sect. \ref{S:Conclusions} we summarize our conclusions. Additional details, benchmarks, and calculations, including a comparison between the exact and approximate HM closures, are included in the appendices.

\section{Maximum-entropy half-moment closures}\label{S:Motivation}

\subsection{Full and half moments}\label{SS:HMidea}

Our goal is to derive a closed system of equations for half moments of the radiation specific intensity, $I_\nu$. This field depends on the photon frequency, $\nu$, the transport direction, $\hat{\boldsymbol{n}}$, and the location in space and time, ($t,\boldsymbol{r}$), where $\boldsymbol{r}$ denotes the spatial position. Our starting point is the radiative transfer equation, which determines the evolution of $I_\nu$ as
\begin{equation}\label{Eq:RTE}
    \frac{1}{c}
    \partial_t I_\nu
    + \hat{\boldsymbol{n}} \cdot \nabla I_\nu = \eta_\nu - \rho \chi_\nu I_\nu\,,
\end{equation}
where $\eta_\nu$ and $\chi_\nu$ are, respectively, the medium's emissivity and total opacity, while $\rho$ is its mass density and $c$ is the speed of light. The total opacity equals the sum of $\kappa_\nu$, the absorption opacity, and $\sigma_\nu$, the scattering opacity. Assuming coherent, isotropic
scattering and thermal emission according to Kirchhoff's law \citep{Mihalas}, we assume an emissivity of the form
\begin{equation}\label{Eq:RTE_emissivity}
    \eta_\nu(T) = \rho \kappa_\nu B_\nu(T) + \rho \sigma_\nu J_\nu\,,
\end{equation}
where $J_\nu$ is the angle-averaged value of $I_\nu$, while $B_\nu(T)=
(2 h \nu^3/c^2)(e^{h \nu/k_\mathrm{B} T}-1)^{-1}$ is the Planck spectral radiance, with $h$ and $k_\mathrm{B}$ the Planck and Boltzmann constants, respectively. Moment approaches rely on an angle integration of this equation multiplied by successive products of $\hat{\boldsymbol{n}}$, resulting in a hierarchy of equations for radiative moments in which the flux of the $n$-th moment is the evolved field in the $(n+1)$-th equation. This procedure needs to be truncated at a finite number of equations in which the highest-order moment is determined as a function of the previous ones via a physically motivated closure relation.

Two-moment methods, such as the M1 closure by \cite{Levermore1984M1}, consist of closed systems of equations for the radiation energy density, its flux, and the radiation pressure tensor, which can be expressed as
\begin{equation} \label{Eq:EFPdefinitions}
\begin{split}
    E &= \frac{1}{c} \int_0^\infty\mathrm{d}\nu 
					 \oint \mathrm{d}\Omega\,\,
  					 I_\nu(t,\boldsymbol{r},\hat{\boldsymbol{n}}) \\
    F^i&= \frac{1}{c} \int_0^\infty\mathrm{d}\nu 
					 \oint \mathrm{d}\Omega\,\,
  					 I_\nu(t,\boldsymbol{r},\hat{\boldsymbol{n}})\, n^i \\
    P^{ij} &= \frac{1}{c} \int_0^\infty\mathrm{d}\nu 
					 \oint \mathrm{d}\Omega\,\,
  					 I_\nu(t,\boldsymbol{r},\hat{\boldsymbol{n}})\, n^i\,n^j\,\,, 	
\end{split}
\end{equation}
respectively. Here we have rescaled the radiation flux by $c$ in such a way that all of these fields are defined in energy density units. 

Instead, here we aim to construct an HM scheme, for which we choose a splitting direction $\hat{\boldsymbol{d}}$ and define $+$ and $-$ half moments by integrating Eq. \eqref{Eq:RTE} separately over the solid angle regions $\mathcal{S}_+$ and $\mathcal{S}_-$ verifying $\hat{\boldsymbol{n}}\cdot \hat{\boldsymbol{d}}>0$ and $<0$, respectively. Specifically, we define
\begin{equation} \label{Eq:EFPdefinitions_HM}
\begin{split}
    E_\pm &= \frac{1}{c} \int_0^\infty\mathrm{d}\nu 
					 \int_{\mathcal{S}_\pm} \mathrm{d}\Omega\,\,
  					 I_\nu(t,\boldsymbol{r},\hat{\boldsymbol{n}}) \\
    F_\pm^i&= \frac{1}{c} \int_0^\infty\mathrm{d}\nu 
					 \int_{\mathcal{S}_\pm}  \mathrm{d}\Omega\,\,
  					 I_\nu(t,\boldsymbol{r},\hat{\boldsymbol{n}})\, n^i \\
    P^{ij}_\pm &= \frac{1}{c} \int_0^\infty\mathrm{d}\nu 
					 \int_{\mathcal{S}_\pm}  \mathrm{d}\Omega\,\,
  					 I_\nu(t,\boldsymbol{r},\hat{\boldsymbol{n}})\, n^i\,n^j\,, 
\end{split}
\end{equation}
and so the full moments are recovered as
\begin{equation}
    \begin{split}
    E &= E_+ + E_- \\
    F^i&= F^i_+ + F^i_- \\
    P^{ij} &= P^{ij}_+ + P^{ij}_-. 	 
\end{split}
\end{equation}
With these definitions, the $+$ and $-$ radiation moments are independent of each other. Thus, the separate integration of Eq. \eqref{Eq:RTE} over $\mathcal{S}_+$ and $\mathcal{S}_-$ yields subsystems for the $+$ and $-$ moments that remain independent in vacuum and are only coupled through the radiation-matter interaction terms. In this work, we do not consider moments beyond the pressure tensors, meaning that we need a closure for $P^{ij}_\pm$ as functions of $(E_\pm,F_\pm)$. 

\subsection{Maximum-entropy closures}\label{SS:maxentr}

In the maximum-entropy\footnote{This is equivalent to minimum-entropy methods \cite[e.g.,][]{Frank2006}, where $h_s$ is defined with a minus sign.} approach, reviewed in \cite{Mueller1998RET}, a closure relation is obtained by finding the distribution function that maximizes the entropy density of the radiation field,
\begin{equation}\label{Eq:entropy}
    s(I_\nu) = \int_0^\infty \mathrm{d}\nu \oint \mathrm{d}\Omega\, h_s(I_\nu)\,,
\end{equation}
under the constraint that $I_\nu$ reproduces a series of moments via defining relations such as Eq. \eqref{Eq:EFPdefinitions} or \eqref{Eq:EFPdefinitions_HM}. 
Here we use the expression for the entropy of a photon gas, with
\begin{equation}\label{Eq:entropy_density}
    h_s(I_\nu) = \frac{2 k_B \nu^2}{c^3}
    \left(
    (\psi+1)\log(\psi+1)-\psi \log \psi
    \right)\,,
\end{equation}
where $\psi=c^2 I_\nu /2 h \nu^3$.
This procedure, which can more generally be applied to classical or degenerate gases, yields the exact same closures as those obtained by enforcing the entropy inequality, ${\partial_t s + \nabla \cdot \Phi_s \geq 0}$ \citep{Dreyer1987,Mueller1998RET}, where $\Phi_s$ is the entropy flux. This provides a clear physical motivation for the entropy maximization approach.

\subsection{The M1 closure}\label{SS:M1closure}

It is illustrative at this point to showcase how the M1 closure can be obtained by maximizing $s$  
under the constraints
$E = \langle I_\nu \rangle$ and $\boldsymbol{F} = \langle I_\nu \hat{\boldsymbol{n}} \rangle$, where $\langle\cdot\rangle$ denotes $\frac{1}{c} \int^\infty_0 \mathrm{d}\nu \oint \mathrm{d}\Omega$. 
To do this, we can apply the method of Lagrange multipliers in calculus of variations. Namely, we can introduce Lagrange multipliers $a$ and $\boldsymbol{b}$ and replace $s$ with the Lagrangian functional
\begin{equation}
    \mathcal{L}=s(I_\nu) + a(\langle I_\nu \rangle - E)
    + \boldsymbol{b} \cdot (\langle I_\nu \hat{\boldsymbol{n}} \rangle - \boldsymbol{F})\,.
\end{equation}
We then zero the variation of $\mathcal{L}$ for small variations of $I_\nu$ around the function that extremizes $s$, resulting in the equation $\partial_{I_\nu}\mathcal{L}=0$, which is equivalent to
\begin{equation}
    \partial_{I_\nu}h_s + a + \boldsymbol{b}\cdot \hat{\boldsymbol{n}} = 0\,.
\end{equation}
After rescaling $a$ and $\boldsymbol{b}$, it is straightforward to obtain that $I_\nu$ is of the form 
\begin{equation}
    I_\nu \propto \frac{\nu^3}{e^{a\nu (1 - \boldsymbol{b}\cdot\hat{\boldsymbol{n}})}-1}\,,
\end{equation}
and therefore
\begin{equation}\label{Eq:intInu}
    \int_0^\infty \mathrm{d}\nu \, I_\nu \propto
    \frac{1}{(1-\boldsymbol{b}\cdot\hat{\boldsymbol{n}})^4}\,.
\end{equation}
By equating $E = \langle I_\nu \rangle$ and $\boldsymbol{F} = \langle I_\nu \hat{\boldsymbol{n}} \rangle$, we can then invert the Lagrange multipliers
as functions of $E$ and $\boldsymbol{F}$, after which the pressure tensor can be obtained as $\mathbb{P} = \langle I_\nu \hat{\boldsymbol{n}} \hat{\boldsymbol{n}} \rangle$. 

The resulting M1 closure can be expressed as
\begin{equation}\label{Eq:ClosureP}
    P^{ij} = D^{ij} E\,,
\end{equation}
where the components of the Eddington tensor $\mathbb{D}$ are defined as follows:
\begin{equation}\label{Eq:EddTensor}
    D^{ij} = \frac{1-\xi}{2} \delta^{ij} +
    \frac{3 \xi - 1}{2} n^i n^j\,.
\end{equation}
As derived by \cite{Levermore1981FLD}, the Eddington factor $\xi$ in this expression has the form
\begin{equation}\label{Eq:ClosureM1}
\xi_\mathrm{M1}(f) = \frac{5-2 \sqrt{4-3 f^2}}{3}\,,
\end{equation}
where $f=||\boldsymbol{F}||/E$ and $n^i=F^i/||\boldsymbol{F}||$, while $\delta^{ij}$ is the Kronecker delta. The diffusion and free-streaming limits, given by ${f \ll 1}$ and $f \to 1$, correspond to $D^{ij} = \delta^{ij}/3$ and $D^{ij} = n^i n^j$, respectively.

\subsection{Half-moment closures for 1D transport}\label{SS:maxentrHM}

The same procedure can be used to find closures for the defined half moments.
In this case, we need to maximize Eq. \eqref{Eq:entropy} for fixed $E_\pm = \langle I_\nu \rangle_{\mathcal{S}_\pm}$ and $\boldsymbol{F}_\pm = \langle I_\nu \hat{\boldsymbol{n}} \rangle_{\mathcal{S}_\pm}$ where the brackets' subscripts indicate the angular region of integration. Since we have twice the number of constraints as before, we also need to duplicate the number of Lagrange multipliers, thereby defining the Lagrangian as
\begin{equation}
\begin{split}
    \mathcal{L} = s(I_\nu) & + a_+ (\langle I_\nu \rangle_{\mathcal{S}_+} - E_+)
    + \boldsymbol{b}_+ \cdot (\langle I_\nu \hat{\boldsymbol{n}} \rangle_{\mathcal{S}_+}  - \boldsymbol{F}_+)\\
    & + a_- (\langle I_\nu \rangle_{\mathcal{S}_-} - E_-)
    + \boldsymbol{b}_- \cdot (\langle I_\nu \hat{\boldsymbol{n}} \rangle_{\mathcal{S}_-}  - \boldsymbol{F}_-)\,.
\end{split}
\end{equation}
Following the same steps as before, we obtain
\begin{equation}
    \partial_{I_\nu}h_s + a_\pm + \boldsymbol{b}_\pm\cdot \hat{\boldsymbol{n}} = 0\,\,\,,\,\,\hat{\boldsymbol{n}}\cdot \hat{\boldsymbol{d}} \gtrless 0\,,
\end{equation}
which results in
\begin{equation}\label{Eq:LagrangeHM}
    \int_0^\infty \mathrm{d}\nu \, I_\nu \propto
    \frac{1}{(1-\boldsymbol{b}_\pm\cdot\hat{\boldsymbol{n}})^4}
    \,\,\,,\,\,\hat{\boldsymbol{n}}\cdot \hat{\boldsymbol{d}} \gtrless 0\,.
\end{equation}
Although this expression is almost identical to Eq. \eqref{Eq:intInu}, its angle integration is substantially more complicated unless $\hat{\boldsymbol{d}}$ and $\boldsymbol{b}_\pm$ can be assumed to be parallel, in which case integration is straightforward in spherical coordinates. Such is the case for slab and spherically symmetric geometries (1D transport).
In these cases, the HM closure by \cite{Dubroca2002} is obtained, where the Eddington factor $\xi$, relating the pressure tensor and energy as $P^{xx}_\pm = \xi E_\pm$, is computed as
\begin{equation}\label{Eq:xiHM}
\xi_\mathrm{HM}(f) = \frac{8 f^2}{1 + 6 f + \sqrt{1 + 12 f (1-f)}}\,,
\end{equation}
where $f=|F_{\pm}|/E_{\pm}$.

\begin{figure}[t!]
\centering
\includegraphics[width=0.95\linewidth]{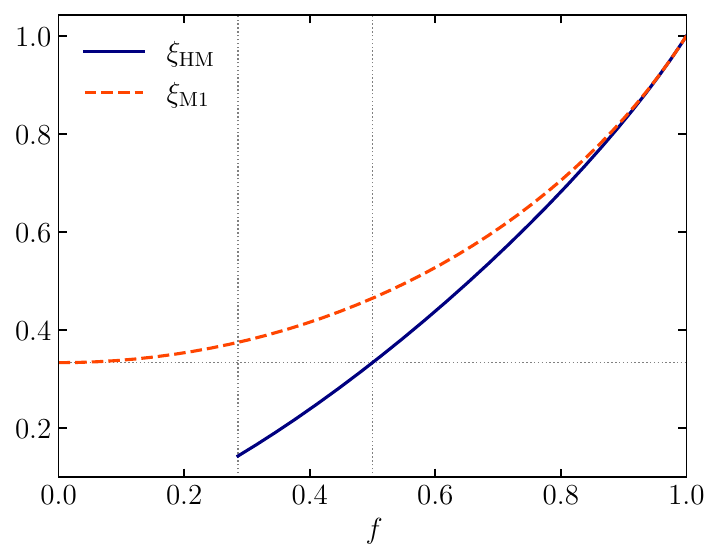}
\caption{
Comparison of the Eddington factor in this work's half-moment closure ($\xi_\mathrm{HM}$) and the M1 closure ($\xi_\mathrm{M1}$) as functions of $f=||\boldsymbol{F}_\pm||/E$ and $||\boldsymbol{F}||/E$, respectively. Vertical dotted lines indicate $f=2/7$ (HM minimum flux) and
$1/2$ (isotropic limit in the HM case), while the horizontal dotted line indicates $\xi=1/3$ (isotropic limit).
}
\label{fig:xi}
\end{figure}

As in the M1 case, this expression tends to $1/3$ in the diffusion regime and $1$ in the free-streaming regime (see Fig. \ref{fig:xi}). However, given that we have integrated over a hemisphere, now the diffusion regime corresponds to $f \approx 1/2$; that is, $F^x_\pm \approx \pm \frac{E_\pm}{2}$, and not to $f=0$. Moreover, $f$ must satisfy
\begin{equation}\label{Eq:f2_7}
    f \geq 2/7\,,
\end{equation}
which results from the exact computation of the half moments from Eq. \eqref{Eq:LagrangeHM}.

\subsection{Half-moment closure used in this work}\label{SS:HM3D}

The HM approach can be immediately generalized to multidimensional transport. Once $\hat{\boldsymbol{d}}$ is fixed, the fluxes defined by Eqs. \eqref{Eq:EFPdefinitions_HM} can have arbitrary directions provided that $\boldsymbol{F}_\pm\cdot \hat{\boldsymbol{d}}\gtrless0$.
In this case, the maximum-entropy HM closure requires a numerical inversion of the Lagrange multiplier in Eq. \eqref{Eq:LagrangeHM}. Instead, here we seek an expression that resembles the exact closure and matches it exactly in the free-streaming and diffusion regimes. With this goal in mind, we define a multidimensional HM closure as follows:
\begin{enumerate}
    \item We assume the Eddington tensor depends on $\xi$ as in Eq. \eqref{Eq:EddTensor}
    for $f\geq 1/2$, with $\hat{\boldsymbol{n}}=\boldsymbol{F}_{\pm}/||\boldsymbol{F}_{\pm}||$.
    \item We compute $\xi$ as in Eq. \eqref{Eq:xiHM}, defining $f=||\boldsymbol{F}_{\pm}||/E_{\pm}$.
    \item In a basis $\{\hat{\boldsymbol{x}},\hat{\boldsymbol{y}},\hat{\boldsymbol{d}}\}$, with $\boldsymbol{\hat{y}}$ perpendicular to $\boldsymbol{F}_{\pm}$, it follows from Eqs. \eqref{Eq:EFPdefinitions_HM} and \eqref{Eq:LagrangeHM} that all components of $\mathbb{D}$ are nonnegative. We enforce this property by flooring the term $(3\xi-1)/2$ in Eq. \eqref{Eq:EddTensor} to zero for $i\neq j$ in such a basis. Expressing this in coordinate-independent form, we compute the Eddington tensor for $f<1/2$ as
    \begin{equation}
    \label{Eq:Df_sm_1_2}
    \begin{split}
        D^{ij} = &\frac{1-\xi}{2}\delta^{ij}
        + \frac{3\xi-1}{2}
        \bigg[
        n^i n^j \\
        & -\big(\hat{\boldsymbol{n}}\cdot\hat{\boldsymbol{d}}\big)
        \big(
        n^i d^j+n^j d^i
        \big)
        + 2 \big(\hat{\boldsymbol{n}}\cdot\hat{\boldsymbol{d}}\big)^2
        d^id^j
        \bigg]\,,
    \end{split}
    \end{equation}
    where $d^i$ are the components of $\hat{\boldsymbol{d}}$. Details leading to this expression can be found in Appendix \ref{A:approximate_exact}.
\end{enumerate}
This closure guarantees that both desired limits are correctly recovered as in the M1 closure, with the difference that now the isotropic limit $D^{ij}=\delta^{ij}/3$ is obtained for $f=1/2$, corresponding to $\boldsymbol{F}_\pm=\pm \frac{E_\pm}{2} \hat{\boldsymbol{d}}$.
Moreover, the exact HM closure, coinciding with that in \cite{Dubroca2002}, is recovered when $\boldsymbol{F}^\pm$ and $\hat{\boldsymbol{d}}$ are parallel.
As is shown in Appendix \ref{A:approximate_exact}, the resulting Eddington tensor closely approximates its functional form in the exact closure with typical differences under $\sim10\%$, vanishing near the mentioned limits.

Both in the exact and approximate HM closures, the radiation fields must verify a few physicality constraints. The first one is the usual limit preventing energy transport faster than the speed of light; namely,
\begin{equation}\label{Eq:physlimit1}
    ||\boldsymbol{F}_\pm||\leq E_\pm\,,
\end{equation}
which stems from Eqs. \eqref{Eq:EFPdefinitions_HM}. The second one is a generalization of Eq. \eqref{Eq:f2_7} given by
\begin{equation}\label{Eq:physlimit2}
    (\boldsymbol{F}_{\pm,\perp})^2 + \left(\frac{7}{2}\right)^2(\boldsymbol{F}_{\pm,\parallel})^2
    \geq E_\pm^2 \,,
\end{equation}
where we have defined $\boldsymbol{F}_{\pm,\parallel} = (\boldsymbol{F}_\pm \cdot \hat{\boldsymbol{d}})\hat{\boldsymbol{d}}$ as the projection of $\boldsymbol{F}_\pm$ onto $\hat{\boldsymbol{d}}$ and $\boldsymbol{F}_{\pm,\perp}=\boldsymbol{F}_\pm-\boldsymbol{F}_{\pm,\parallel}$. 
We obtained this relation from a numerical computation of the half moments using Eq. \eqref{Eq:LagrangeHM} (Appendix \ref{A:approximate_exact}). Lastly, the fluxes must verify $\boldsymbol{F}_+\cdot \hat{\boldsymbol{d}} >0$, $\boldsymbol{F}_-\cdot \hat{\boldsymbol{d}} <0$, which stems from their definition. 
Although these constraints are naturally enforced in the diffusion regime, this is not necessarily the case for free-streaming transport. We investigate the consequences of enforcing them in Sections \ref{SS:Shadow} and \ref{S:Disks}.

Computing the pressure tensor in these expressions is significantly less expensive than in the exact maximum-entropy HM closure since it can be done explicitly, as in the M1 closure. This makes the approximate HM closure ideal for implementation in numerical solvers for hyperbolic systems, such as those summarized in Section \ref{S:NumericalImpl}.

\subsection{Multigroup extension}\label{SS:MultigroupHM}

We now turn to a generalization to multigroup moments. We divide the frequency domain in $N_g$ intervals or frequency groups, $[\nu^\mathrm{min}_{\ell},\nu^\mathrm{max}_{\ell}]$, where $\ell=1,\ldots,N_g$ indicates the group index (this notation is kept throughout this work). For each group, we define half moments as
\begin{equation} \label{Eq:EFPmultigroup}
\begin{split}
    E_{\ell,\pm} &= \frac{1}{c} \int^{\nu^\mathrm{max}_{\ell}}_{\nu^\mathrm{min}_{\ell}}\mathrm{d}\nu 
					 \int_{S_\pm} \mathrm{d}\Omega\,\,
  					 I_\nu(t,\boldsymbol{r},\hat{\boldsymbol{n}}) \\
    F_{\ell,\pm}^i&= \frac{1}{c}  \int^{\nu^\mathrm{max}_{\ell}}_{\nu^\mathrm{min}_{\ell}}\mathrm{d}\nu 
					 \int_{S_\pm}  \mathrm{d}\Omega\,\,
  					 I_\nu(t,\boldsymbol{r},\hat{\boldsymbol{n}})\, n^i \\
    P_{\ell,\pm}^{ij} &= \frac{1}{c}  \int^{\nu^\mathrm{max}_{\ell}}_{\nu^\mathrm{min}_{\ell}}\mathrm{d}\nu 
					 \int_{S_\pm}  \mathrm{d}\Omega\,\,
  					 I_\nu(t,\boldsymbol{r},\hat{\boldsymbol{n}})\, n^i\,n^j\,, 
\end{split}
\end{equation}
in such a way that the total moments are recovered by summing over groups and signs as
\begin{equation}
    E = \sum_{\substack{\ell \\ s=+,-}}E_{\ell,s}\,,\,\,\,\,\,
    \boldsymbol{F} = \sum_{\substack{\ell \\ s=+,-}}\boldsymbol{F}_{\ell,s}\,,\,\,\,\,\,
    \mathbb{P} = \sum_{\substack{\ell \\ s=+,-}}\mathbb{P}_{\ell,s}\,.
\end{equation}

To obtain a closure for the defined half moments, we would need to repeat the entropy maximization process and integrate the radiative intensity within each frequency group to obtain similar expressions to Eqs. \eqref{Eq:intInu} and \eqref{Eq:LagrangeHM}, using different $\pm$ Lagrange multipliers per group. Unfortunately, as for the multigroup M1 closure, this frequency integration cannot be achieved analytically for finite integration intervals, which means that the resulting Eddington factor can only be obtained either via numerical inversion of the Lagrange multipliers in each group.
This approach has been followed in \cite{Turpault2002,Turpault2003thesis,Turpault2005,Ripoll2008} for the M1 closure and in \cite{Turpault2004} for the 1D HM closure.

Instead, here we follow a simplified approach often used in multigroup M1 methods \citep{Vaytet2011,Rosdahl2013,AnninosFragile2020}, which consists of using the same closure for each group as in the frequency-integrated case. For each group, we compute $\mathbb{P}_{\ell,\pm}$ in the same way as in Section \ref{SS:HM3D}, this time using $f=||\boldsymbol{F}_{\ell,\pm}||/E_{\ell,\pm}$ and $\hat{\boldsymbol{n}}=\boldsymbol{F}_{\ell,\pm}/||\boldsymbol{F}_{\ell,\pm}||$ to obtain the Eddington factors via Eq. \eqref{Eq:xiHM}. 
This approach is justified by the property that the Eddington tensor must still have the form of Eq. \eqref{Eq:EddTensor}
in the diffusion and free-streaming limits.
We adopt this method as a first approximation and defer a precise comparison with the exact maximum-entropy closure to future studies.
\section{Half-moment and M1 radiation MHD}\label{S:HalfMoment}
\subsection{Multigroup half-moment radiative transfer}\label{SS:HMequations}

Here we summarize the evolution equations used in this work, beginning with HM radiative transfer in the general multigroup case. These can be obtained by integrating Eq. \eqref{Eq:RTE} in frequency and solid angle in the hemispheres $\mathcal{S}_\pm$, resulting in the following system:
\begin{equation}
\begin{split}\label{Eq:HMRT_multigroup}
    \frac{1}{\hat{c}}\partial_t E_{\ell,\pm} + \nabla \cdot \boldsymbol{F}_{\ell,\pm}  &= - G^0_{\ell,\pm}  \\
    \frac{1}{\hat{c}} \partial_t \boldsymbol{F}_{\ell,\pm}  + \nabla \cdot \mathbb{P}_{\ell,\pm}  &=
    - \boldsymbol{G}_{\ell,\pm}  
    \,,
\end{split}
\end{equation}
where the components of $\mathbb{P}_{\ell,\pm} $ are obtained as functions of $(E_{\ell,\pm} ,\boldsymbol{F}_{\ell,\pm} )$ as specified in Section \ref{SS:MultigroupHM}. To increase the maximum time step allowed by the explicit integration of transport terms, here we have replaced $c$ with the reduced speed of light $\hat{c}<c$. 

When coupled to HD or MHD, the energy and momentum equations read
\begin{equation}\label{Eq:HMRT_multigroup_MHD}
\begin{split}
    \partial_t \mathcal{E} + S^\mathcal{E}_\mathrm{HD/MHD} &= c \sum_\ell (G^0_{\ell,+}+G^0_{\ell,-})
    - \nabla \cdot \boldsymbol{F}_\mathrm{Irr} \\
    \partial_t \boldsymbol{m} + \boldsymbol{S}^{\boldsymbol{m}}_\mathrm{HD/MHD} 
    &= \sum_\ell (\boldsymbol{G}_{\ell,+}+\boldsymbol{G}_{\ell,-}) \,,
\end{split}
\end{equation}
where $\mathcal{E}$ and $\boldsymbol{m}$ are the gas total energy and momentum densities, respectively.
Here we have only explicitly written terms corresponding to radiative processes, summarizing the rest (e.g., advection and pressure terms) as $S^\mathcal{E}_\mathrm{HD/MHD}$ and $\boldsymbol{S}^{\boldsymbol{m}}_\mathrm{HD/MHD}$. We have also added a source term for external irradiation sources producing a flux $\boldsymbol{F}_\mathrm{Irr}$, which will be used when modeling irradiated disks in Section \eqref{S:Disks}.

Radiation-matter interaction terms, obtained by integrating the right-hand side of Eq. \eqref{Eq:RTE} (Appendix \ref{A:sourceterms}), are computed for constant opacities as
\begin{subequations}\label{Eq:HMSourceTerms_grey}
\begin{align}
    G^0_{\ell,\pm}  &= \rho \kappa_\ell \left(E_{\ell,\pm}  - \frac{B_\ell(T)}{2}\right)
    + \rho \sigma_\ell \left(
    E_{\ell,\pm}  - \frac{E_\ell}{2} 
    \right) \label{Eq:HMSourceTerms_grey_1}
    \\
    \boldsymbol{G}_{\ell,\pm}  &= 
    \rho \kappa_\ell \left(
     \boldsymbol{F}_{\ell,\pm} 
     \mp
    \frac{B_\ell(T)}{4} \, \hat{\boldsymbol{d}}
    \right)
    +
    \rho \sigma_\ell \left(
     \boldsymbol{F}_{\ell,\pm} 
     \mp
    \frac{E_\ell}{4} \, \hat{\boldsymbol{d}}
    \right)\,, \label{Eq:HMSourceTerms_grey_2}
\end{align}
\end{subequations}
where $\rho$ is the matter density and $\kappa_\ell$, $\sigma_\ell$, and $\chi_\ell$ are the group absorption, scattering, and total opacities, respectively, while
${E_\ell=E_{_\ell,+}+E_{_\ell,-}}$ and 
\begin{equation}
    B_\ell(T) = \frac{4\pi}{c}  \int^{\nu^\mathrm{max}_{\ell}}_{\nu^\mathrm{min}_{\ell}}B_\nu(T)\,\mathrm{d}\nu 
\end{equation}
is the integrated Planck energy density, where $T$ is the temperature. In the single-group case with $[\nu^\mathrm{min},\nu^\mathrm{max})=[0,\infty)$, this function becomes $B(T)=a_R T^4$, where $a_R$ is the radiation constant.

For frequency-dependent opacities, the coefficients in Eq. \eqref{Eq:HMSourceTerms_grey_1} are replaced with their Planck-averaged values,
\begin{equation}
    \kappa_{P\ell} = \frac{
    \int^{\nu^\mathrm{max}_{\ell}}_{\nu^\mathrm{min}_{\ell}}
    \kappa_\nu
    \, B_\nu(T)\, \mathrm{d}\nu
    }{
    \int^{\nu^\mathrm{max}_{\ell}}_{\nu^\mathrm{min}_{\ell}}
    \, B_\nu(T)\, \mathrm{d}\nu
    }\,,\,\,\,
    \sigma_{P\ell} = \frac{
    \int^{\nu^\mathrm{max}_{\ell}}_{\nu^\mathrm{min}_{\ell}}
    \sigma_\nu
    \, B_\nu(T)\, \mathrm{d}\nu
    }{
    \int^{\nu^\mathrm{max}_{\ell}}_{\nu^\mathrm{min}_{\ell}}
    \, B_\nu(T)\, \mathrm{d}\nu
    }\,,
\end{equation}
while Eq. \eqref{Eq:HMSourceTerms_grey_2} is replaced with
\begin{equation}\label{Eq:GpmAbsorption}
        \boldsymbol{G_{\ell,\pm}}=
        \rho \kappa_{R\ell}
        \left(
        \boldsymbol{F}_{\ell,\pm}
        \mp \frac{B_\ell(T)}{4}\hat{\boldsymbol{d}}
        \right)
\end{equation}
for pure absorption opacities and
\begin{equation}\label{Eq:GpmScattering}
        \boldsymbol{G_{\ell,\pm}}=
        \rho \sigma_{R\ell}
        \left(
        \boldsymbol{F}_{\ell,\pm}
        \mp \frac{E_\ell}{4}\hat{\boldsymbol{d}}
        \right)
\end{equation}
for pure scattering. Here the subscript $R$ denotes Rosseland averaging, defined as
\begin{equation}
    \kappa_{R\ell}^{-1}= \frac{
    \int^{\nu^\mathrm{max}_{\ell}}_{\nu^\mathrm{min}_{\ell}}
    \kappa_\nu^{-1}
    \, \frac{\partial B_\nu(T)}{\partial T}\, \mathrm{d}\nu
    }{
    \int^{\nu^\mathrm{max}_{\ell}}_{\nu^\mathrm{min}_{\ell}}
    \,  \frac{\partial B_\nu(T)}{\partial T}\, \mathrm{d}\nu
    }\,,\,\,\,
    \sigma_{R\ell}^{-1}= \frac{
    \int^{\nu^\mathrm{max}_{\ell}}_{\nu^\mathrm{min}_{\ell}}
    \sigma_\nu^{-1}
    \, \frac{\partial B_\nu(T)}{\partial T}\, \mathrm{d}\nu
    }{
    \int^{\nu^\mathrm{max}_{\ell}}_{\nu^\mathrm{min}_{\ell}}
    \,  \frac{\partial B_\nu(T)}{\partial T}\, \mathrm{d}\nu
    }\,.
\end{equation}
The case of frequency-dependent combined absorption and scattering is addressed in Appendix \ref{A:sourceterms}.

\subsection{Multigroup M1 radiative transfer}\label{SS:MultigroupM1}

We now introduce the multigroup M1 equations solved in this work, which generalize the single-group system considered in \cite{MelonFuksman2021}. These are expressed as
\begin{equation}
\begin{split}\label{Eq:M1RT}
    \frac{1}{\hat{c}}\partial_t E_\ell + \nabla \cdot \boldsymbol{F}_\ell &= - G^0_{\ell}  \\
    \frac{1}{\hat{c}} \partial_t \boldsymbol{F}_\ell + \nabla \cdot \mathbb{P}_\ell &=
    - \boldsymbol{G}_\ell 
    \,,
\end{split}
\end{equation}
where the angle-averaged moments are defined for each group analogously to Eq. \eqref{Eq:EFPmultigroup}, this time using closed solid angle integrations. As in our HM closure and the M1 methods by \cite{Vaytet2011,Rosdahl2013,AnninosFragile2020}, we compute $\mathbb{P}_\ell(E_\ell,\boldsymbol{F}_\ell)$ using the same M1 closure as in the frequency-integrated case (Section \ref{SS:M1closure}).
The radiation fields are then coupled with HD and MHD as follows:
\begin{equation}\label{Eq:M1RT_multigroup_MHD}
\begin{split}
    \partial_t \mathcal{E} + S^\mathcal{E}_\mathrm{HD/MHD} &= c \sum_\ell G^0_\ell
    - \nabla \cdot \boldsymbol{F}_\mathrm{Irr} \\
    \partial_t \boldsymbol{m} + \boldsymbol{S}^{\boldsymbol{m}}_\mathrm{HD/MHD} &= \sum_\ell \boldsymbol{G}_\ell \,,
\end{split}
\end{equation}
where the source terms are computed as 
\begin{equation}
\begin{split}
    G^0_{\ell} &= \rho \kappa_{P\ell} \left(E_{\ell} - B_\ell(T)\right)
    \\
    \boldsymbol{G}_{\ell} &= \rho \chi_{R\ell} 
     \boldsymbol{F}_{\ell}\,.
\end{split}
\end{equation}

\section{Numerical implementation}\label{S:NumericalImpl}

To solve Equations \eqref{Eq:HMRT_multigroup} and \eqref{Eq:HMRT_multigroup_MHD}, we implemented a numerical scheme that extends the gray M1 radiative transfer code by \cite{MelonFuksman2021}, which is freely distributed\footnote{\url{https://plutocode.ph.unito.it/}} as part of PLUTO 4.4-patch3. Our solution strategy is based on an explicit integration of advection terms via Godunov solvers and a locally implicit integration of the potentially stiff radiation-matter interaction terms. Operator splitting is used to integrate the transport equation of the radiation and MHD fields separately, each with their corresponding maximum time steps allowed by the CFL stability criterion. The integration of radiation transport and interaction terms is carried out simultaneously using the same IMEX schemes implemented for the M1 closure in \cite{MelonFuksman2019}. While here we only describe our implementation of the multigroup HM method, we use analogous methods to integrate the equations of both single-group and multigroup HM and M1 radiative transfer.

\subsection{Explicit step}\label{SS:TransportTerms}

The explicit integration of radiation transport terms involves the solution of $2 N_g$ independent systems of hyperbolic PDEs of the form
\begin{equation}\label{Eq:hyperbolicHM}
    \partial_t \mathcal{U}_{\ell,s} + \nabla \cdot \bm{\mathcal{F}}_{\ell,s} = 0
\end{equation}
corresponding to the left-hand side of Eqs. \eqref{Eq:HMRT_multigroup}, where $\mathcal{U}_{\ell,s}=(E_{\ell,s},\boldsymbol{F}_{\ell,s})$, with $s=+$ or $-$. We solve these systems independently using the same reconstruct-solve-update strategy as in PLUTO, following a finite volume approach in which the discretized values of $\mathcal{U}_{\ell,s}$ are interpreted as cell averages. In this implementation, $\hat{\boldsymbol{d}}$ must be parallel to one of the coordinate axes in the chosen grid geometry; that is, it can only align with a Cartesian, cylindrical, or spherical unit vector.

To compute the fluxes, a set $\mathcal{V}_{\ell,s}$ of primitive fields mapped from $\mathcal{U}_{\ell,s}$ must be interpolated into the inter-cell interfaces. Here we chose $\mathcal{V}_{\ell,s}=\mathcal{U}_{\ell,s}$. We also explored definitions such as $\mathcal{V}_{\ell,\pm}=\left(E_{\ell,\pm},\boldsymbol{F}_{\ell,\pm}\mp (E_{\ell,\pm}/2)\hat{\boldsymbol{d}}\right)$ to properly compute the flux in diffusion problems where $||\boldsymbol{F}_{\ell}||\ll E_\ell$ but
$\boldsymbol{F}_{\ell,\pm}\approx \pm(E_{\ell,\pm}/2) \hat{\boldsymbol{d}}$. However, we found that the former definition of $\mathcal{V}_{\ell,s}$ suffices to reproduce the correct diffusion flux in all cases we considered.

The reconstructed fields are then used to compute inter-cell fluxes via Riemann solvers. We implemented a Lax–Friedrichs (LF) solver and a Harten–Lax–van Leer (HLL) solver identical in form to those in \cite{MelonFuksman2019}. These solvers require knowledge of the maximum and minimum signal speeds along each direction, which we compute as the eigenvalues of the Jacobian matrices of Eq. \eqref{Eq:hyperbolicHM}. These can be explicitly obtained as functions of the radiation fields, as detailed in Appendix \ref{A:SignalSpeeds}.

The obtained fluxes are finally used to update the cell-averaged $\mathcal{U}_{\ell,s}$, while signal speeds are used to update the time step according to the CFL condition. The latter step motivates our use of a reduced speed of light, which is not a necessary feature of the HM closure.

\subsection{Implicit step}\label{SS:SourceTerms}

The implicit step handles the integration of radiation-matter interaction terms. By adding Equations \eqref{Eq:HMRT_multigroup} and \eqref{Eq:HMRT_multigroup_MHD} zeroing all transport and MHD terms, we can see that this step conserves the following modified total energy and momentum densities:
\begin{equation}\label{Eq:Etot_mtot}
\begin{split}
    E_\mathrm{tot}&=\mathcal{E} + \frac{c}{\hat{c}}\sum_{\ell,s} E_{\ell,s}\\
    \boldsymbol{m}_\mathrm{tot}&=\boldsymbol{m}+\frac{1}{\hat{c}}\sum_{\ell,s}\boldsymbol{F}_{\ell,s}\,,
\end{split}
\end{equation}
which coincide with the total energy and momentum only if $\hat{c}=c$. As extensively tested in various physical contexts \citep[e.g.,][]{Skinner2013,MelonFuksman2021,MelonFuksman2024partI,MelonFuksman2022}, these modified conservation laws do not alter the solutions for large enough $\hat{c}$. If external irradiation is included, $-\nabla\cdot \boldsymbol{F}_\mathrm{Irr}$ is simply added to $E_\mathrm{tot}$ (see Appendix \ref{A:ImplicitMethod}), while if
magnetic fields are included, the magnetic energy density $\boldsymbol{B}^2/2$ is subtracted from $\mathcal{E}$ as it does not intervene in the radiative energy exchange.

We use these conservation laws to simplify the equations solved during the implicit step, which become
\begin{equation}
\begin{split}\label{Eq:HMRT_implicit}
    \partial_t E_{\ell,\pm}  &= - \hat{c}\, G^0_{\ell,\pm} \\
    \partial_t \boldsymbol{F}_{\ell,\pm} &=
    - \hat{c}\, \boldsymbol{G}_{\ell,\pm} 
    \,,
\end{split}
\end{equation}
together with the conservation equtions of $E_\mathrm{tot}$ and $\boldsymbol{m}_\mathrm{tot}$. Unlike in the explicit step, all of these equations must be solved simultaneously due to the coupling introduced by the interaction terms. To speed up this process, we use the fact that momentum and flux variations resulting from radiative processes are much smaller than energy variations, allowing us to update all energies via either Newton or fixed-point methods while updating $\boldsymbol{F}_{\ell,\pm}$ in a fixed-point fashion. We describe these methods in Appendix \ref{A:ImplicitMethod}.

\section{Numerical benchmarks}\label{S:Tests}

In this section, we test our implementation of HM radiative transfer in single-group problems designed to evaluate the main features of our methods. All tests were performed using the IMEX1 method in \cite{MelonFuksman2019} to integrate the radiation fields, switching off HD advection. Unless otherwise stated, zero-gradient conditions were imposed on all fields, and fluxes were computed using the HLL solver and linear reconstruction
with the second-order limiter by \cite{VanLeer1974}. Additional tests involving absorption and scattering in the diffusion and free-streaming limits are presented in Appendix \ref{A:AdditionalTests}.

\subsection{Freely streaming beams}\label{SS:FreeStreaming}

As a starting point, we tested the multidimensional transport scheme in the free-streaming regime. We studied the propagation of freely streaming beams oblique to the grid to evaluate their spread in the most diffusive case possible \citep{Gonzalez2007}. We also performed crossing beam tests to demonstrate how the HM approach prevents the artificial interaction between converging beams in specific directions.

\begin{figure}[t!]
\centering
\includegraphics[width=\linewidth]{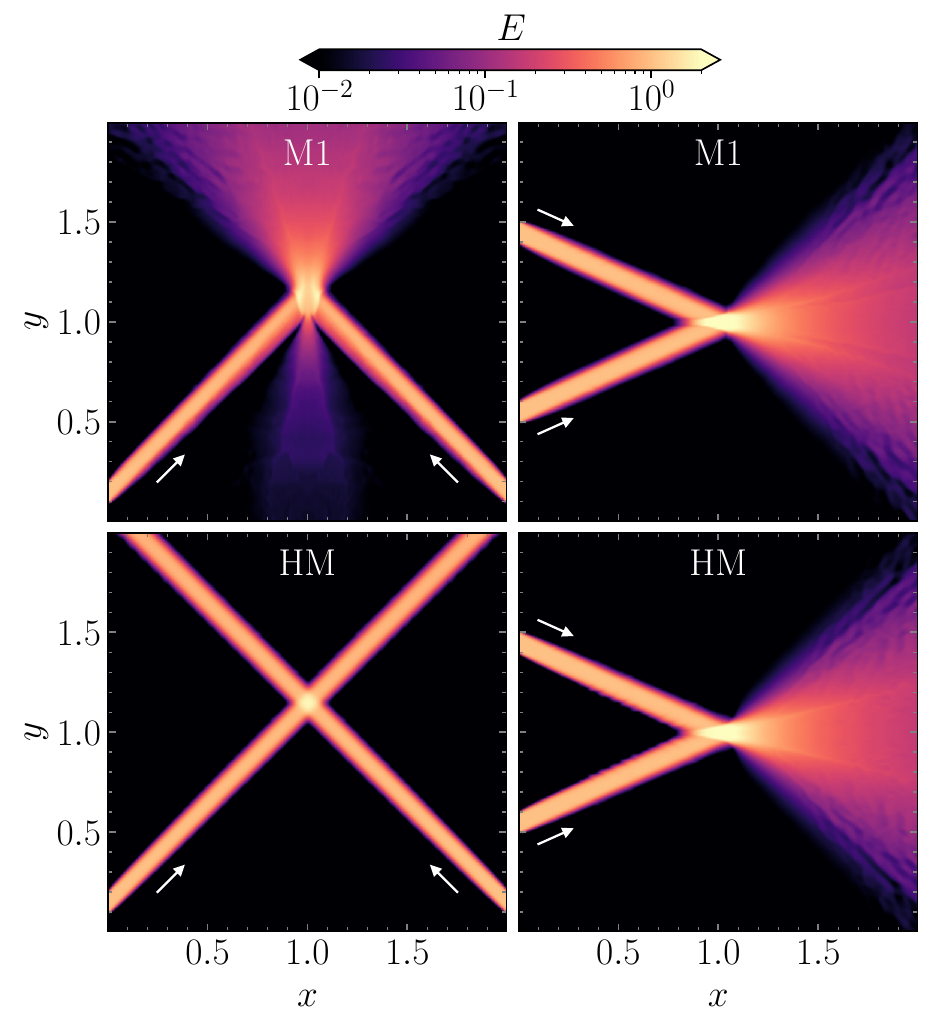}
\caption{Crossing beams produced with the M1 and HM closures. This test highlights how the HM closure removes the artificial interaction between crossing fluxes in selected directions. In the HM case, $\hat{\boldsymbol{d}}=\hat{\boldsymbol{x}}$ has been chosen as the splitting direction. Arrows indicate the directions of the injected beams.}
\label{fig:crossing_beams}
\end{figure}

We defined the computational domain as $[0,2]\times[0,2]$ in 2D Cartesian coordinates, with $300$ cells in each direction. Fixing $\hat{\boldsymbol{d}}=\hat{\boldsymbol{x}}$ as the splitting direction, we initially set $E_+=E_-=10^{-8}$ in the isotropic transport regime; namely, $\boldsymbol{F}_\pm=\pm (E_\pm/2) \hat{\boldsymbol{d}}$. At the left and right boundaries, we injected two beams in the free-streaming regime with $||\boldsymbol{F}_\pm||=E_\pm=1$ as follows:
\begin{equation}
\begin{alignedat}{2}
    &\quad\boldsymbol{F}_+ = (1/\sqrt{2},1/\sqrt{2}) \quad &\text{if $x=0$, $0.1\leq y \leq 0.2$}\,,\\
    &\quad\boldsymbol{F}_- = (-1/\sqrt{2},1/\sqrt{2}) \quad &\text{if $x=2$, $0.1\leq y \leq 0.2$}\,.
\end{alignedat}
\end{equation}
The produced beams, shown in the bottom left panel of Fig. \ref{fig:crossing_beams}, are correctly advected at the speed of light maintaining their original direction. Trivially, the beams cross each other without interacting, as they correspond to separate, non-interacting fields. 
This is always the case when the beams have fluxes with opposite signs along the splitting direction. Conversely, if a third beam converging at the same point were introduced, it would inevitably lead to beam-crossing interaction, as its flux would necessarily have the same sign along the splitting direction as one of the other two beams.

In contrast, we also considered a case in which the crossing beams propagate with positive $x$ velocity. We injected these beams at the left boundary with $||\boldsymbol{F}_\pm||=E_\pm=1$, as
\begin{equation}
  \quad\boldsymbol{F}_+ =\begin{cases}
    (\sqrt{5/6},1/\sqrt{6}) & \text{if $0.5\leq y \leq 0.6$}\\
    (\sqrt{5/6},-1/\sqrt{6}) & \text{if $1.4\leq y \leq 1.5$}
  \end{cases}\,.
\end{equation}
In this case, as with the M1 closure, the beams artificially interact and merge at their crossing point (Fig. \ref{fig:crossing_beams}, right panel), as they are both described by the same field ($E_+$ in this case).

For comparison, we repeated both tests with the M1 closure.
The resulting energy distributions are shown in Fig. \ref{fig:crossing_beams}. The M1 closure always produces an interaction between crossing beams, while the HM closure does so only when both beams have positive $x$ flux. Due to the chosen beam angles, the upward-transported M1 beams produce an overpressurized region at their colliding point that emits beams with both positive and negative $y$ flux. Before any interaction is produced, both HM and M1 beams propagate across the domain, maintaining their initial directions with similar numerical spread. In both cases, numerical broadening decreases when the resolution is doubled. The merged beams in the right panels are similar for both closures because these coincide in the free-streaming limit. In both cases, sharp gradients, combined with grid noise, produce ripples in the energy distribution as radiation propagates away from the convergence point.

\subsection{Shadows}\label{SS:Shadow}

\begin{figure}[t!]
\centering
\includegraphics[width=\linewidth]{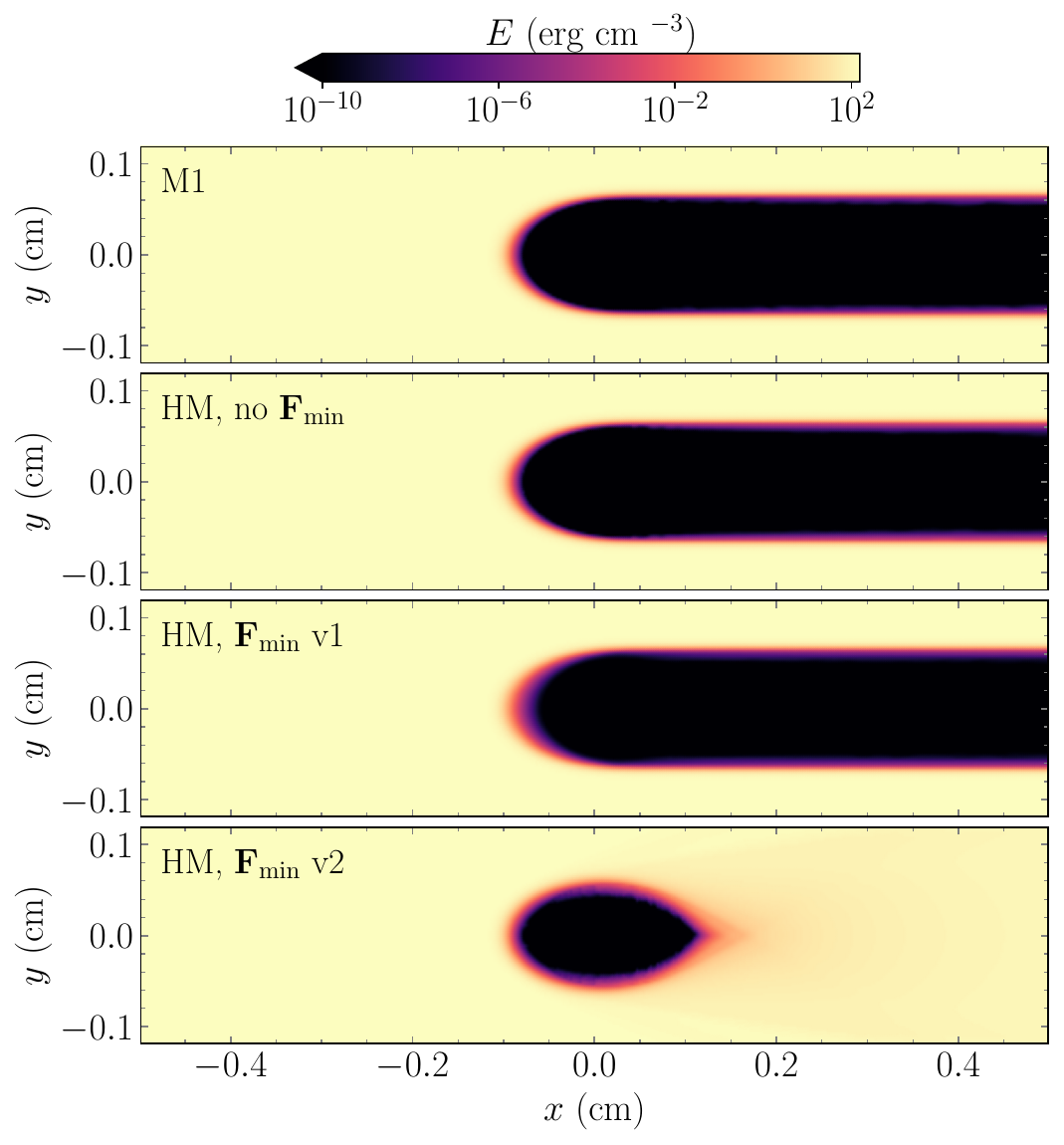}
\caption{Total energy density in the horizontal shadow test obtained with the M1 and HM closures. For the HM closure, distributions are shown for the cases where Eq. \eqref{Eq:physlimit2} is not enforced, enforced via Eq. \eqref{Eq:Fminv1}, and enforced via Eq. \eqref{Eq:Fminv2}, in the second, third, and fourth panels from the top, respectively.
}
\label{fig:shadow}
\end{figure}

The ability to produce shadows is a key feature of the M1 closure, absent in the more diffusive FLD method \citep{Hayes2003}. We tested this property in our HM code by reproducing a shadow test akin to those in \cite{Hayes2003} and \cite{Gonzalez2007}, which also serves to consider a case of transition between the diffusion and free-streaming regimes.

We defined a 2D Cartesian domain as ${[-0.5,0.5]\,\mathrm{cm}\times[-0.12,0.12]\,\mathrm{cm}}$, with a resolution of ${400\times 96}$ cells. We set the density such that it transitioned between a minimum value of $\rho_0=10^{-5}$ g cm$^{-3}$ and a maximum value $\rho_1=10^2$ g cm$^{-3}$ with the functional form
\begin{equation}
    \rho(x,y)=\rho_0 + \frac{\rho_1-\rho_0}{1+e^{\Delta}}\,,
\end{equation}
where $\Delta=10\left[(x/x_0)^2+(y/y_0)^2-1\right]$, with $(x_0,y_0)=(0.1,0.06)$ cm. This defines an elliptic overdense region at the center of the domain, acting as an obstacle for the incoming radiation. To this end, we defined a uniform absorption opacity $\kappa=10$ cm$^2$ $g^{-1}$, resulting in a maximum horizontal optical depth of $\sim 200$ across the obstacle and $\sim 10^{-4}$ through the background medium.

With the system initially in local thermal equilibrium (LTE) at $T=10$ K, we injected a freely streaming radiation flux at the left boundary as $\boldsymbol{F}=E_1\, \hat{\boldsymbol{x}}$, with $E_1= 2 a_R T_1^4$, where $T_1=10^4$ K. This parameterization of $E_1$ is unrelated to the actual temperature distribution of the injected photons, which is undefined.

\begin{figure}[t!]
\centering
\includegraphics[width=\linewidth]{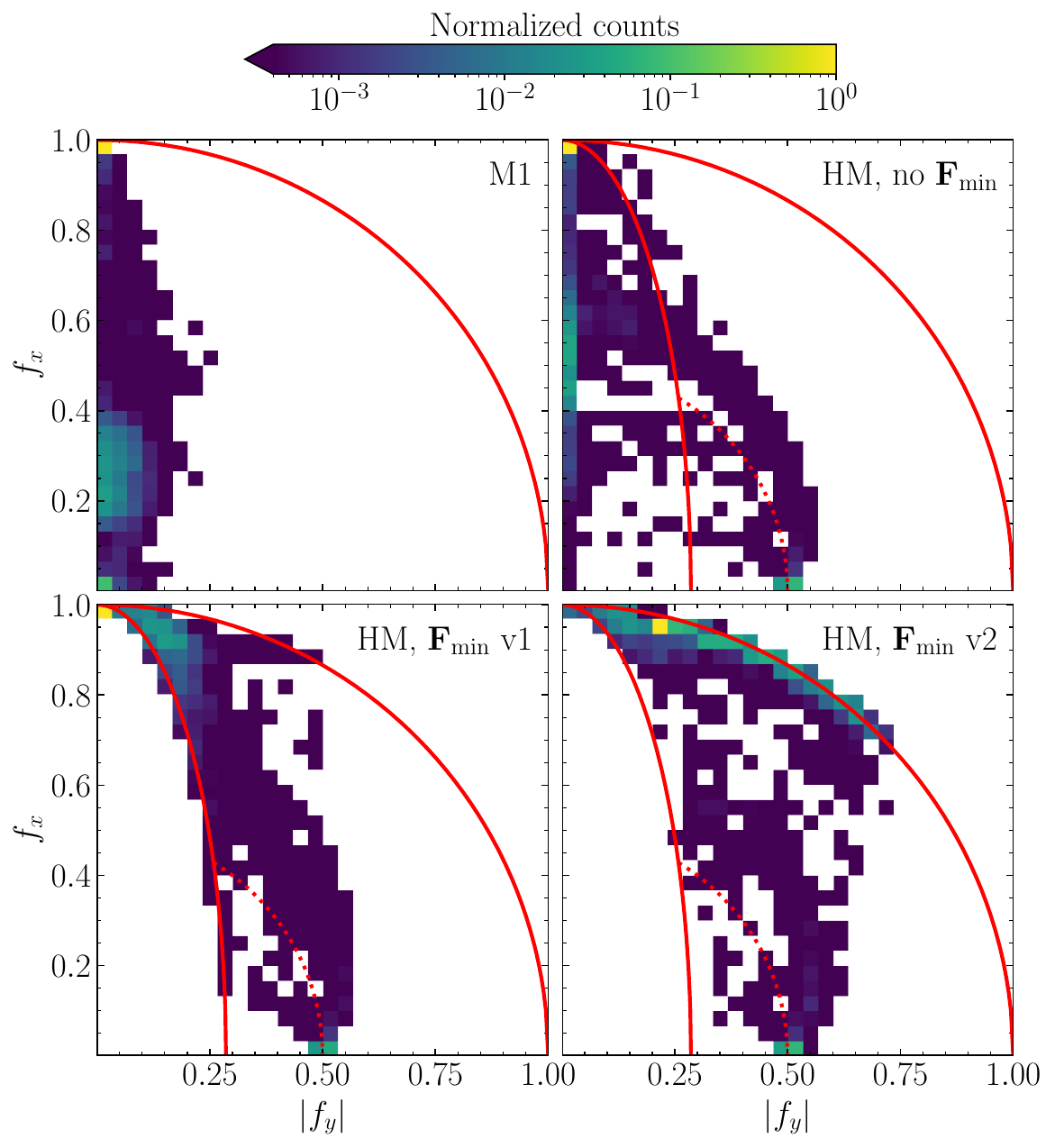}
\caption{Frequency of $(|f_y|,f_x)$ values in each computational cell in the shadow tests shown in Fig. \ref{fig:shadow}, normalized to a maximum of $1$. The reduced fluxes are defined as $(f_x,f_y)=\boldsymbol{F}/E$ and $\boldsymbol{F}_\pm/E_\pm$ in the M1 and HM closures, respectively. Solid lines indicate the limits given by Eqs. \eqref{Eq:physlimit1} and \eqref{Eq:physlimit2}, while dotted lines correspond to $f=1/2$.}
\label{fig:shadow_histograms}
\end{figure}

We ran this simulation with the M1 and HM closures, switching off the update of the density and internal energy in both cases to focus solely on radiation transport. In the HM case, we chose $\hat{\boldsymbol{d}}=\hat{\boldsymbol{y}}$ as the splitting direction, considering equal $+$ and $-$ fluxes. As is shown in the top two panels of Fig. \ref{fig:shadow}, a shadow is correctly formed in both cases behind the obstacle, and the energy distributions in both simulations are almost indistinguishable from each other. This is because, except for the shadowed region, radiation is either in the free streaming regime or close to LTE in the diffusion regime, and the Eddington tensor in both closures coincides in both regimes. In the HM case, this test also shows that radiation is correctly transported perpendicularly to the splitting direction $\hat{\boldsymbol{d}}$.

We also used this test to evaluate the effect of enforcing the minimum-flux constraint given by Eq. \eqref{Eq:physlimit2}. After each IMEX integration, we achieve this by modifying the flux in one of the following two ways wherever $\boldsymbol{F}_\pm$ does not satisfy the condition:
\begin{subequations}
\begin{align}
    \label{Eq:Fminv1}
    &\boldsymbol{F}_\pm \rightarrow \alpha \boldsymbol{F}_\pm \,\,\,\, &\mathrm{(\boldsymbol{F}_\mathrm{min}\,v1)}&\\
    \label{Eq:Fminv2}
    &\boldsymbol{F}_{\pm,||} \rightarrow \pm (E_\pm/2)\hat{\boldsymbol{d}} \,\,\,\, &\mathrm{(\boldsymbol{F}_\mathrm{min}\,v2)}&\,,
\end{align}
\end{subequations}
where $\alpha\in\mathbb{R}_{>0}$ is the minimum number that ensures the constraint is satisfied. 

In Fig. \ref{fig:shadow}, we show a comparison of the energy distributions obtained when this limiting is not enforced versus when each of these transformations is applied. When using Eq. \eqref{Eq:Fminv1}, the shadow looks almost the same as before, but the transition to the diffusion regime where radiation encounters the obstacle is smoother in the radial direction. This is because, in that region, interaction with matter preferentially reduces $F^x_\pm$ as radiation propagates through the obstacle. We can see this in both the M1 and HM cases in the $(|f_y|,f_x)$ histograms shown in Fig. \ref{fig:shadow_histograms} when the constraint is not applied. This reduction is not consistent with Eq. \eqref{Eq:physlimit2}, and so the flux is modified in that region. Once in the diffusion regime, with $f\sim1/2$, such a flux limiting is no longer required.

Conversely, only increasing $|F^y_{\pm}|$ as in Eq. \eqref{Eq:Fminv2} leads to a similar transition in the $x$ direction as before, but smooths out the transition in the $y$ direction. Moreover, in this case, the shadow is lost because the method favors transport parallel to $\hat{\boldsymbol{d}}$. We conclude that care must be taken in general when enforcing Eq. \eqref{Eq:physlimit2}. While this is generally not necessary, it can be useful in beam-crossing scenarios that lead to deviations from this constraint (Section \ref{S:Disks}).

\begin{figure}[t!]
\centering
\includegraphics[width=\linewidth]{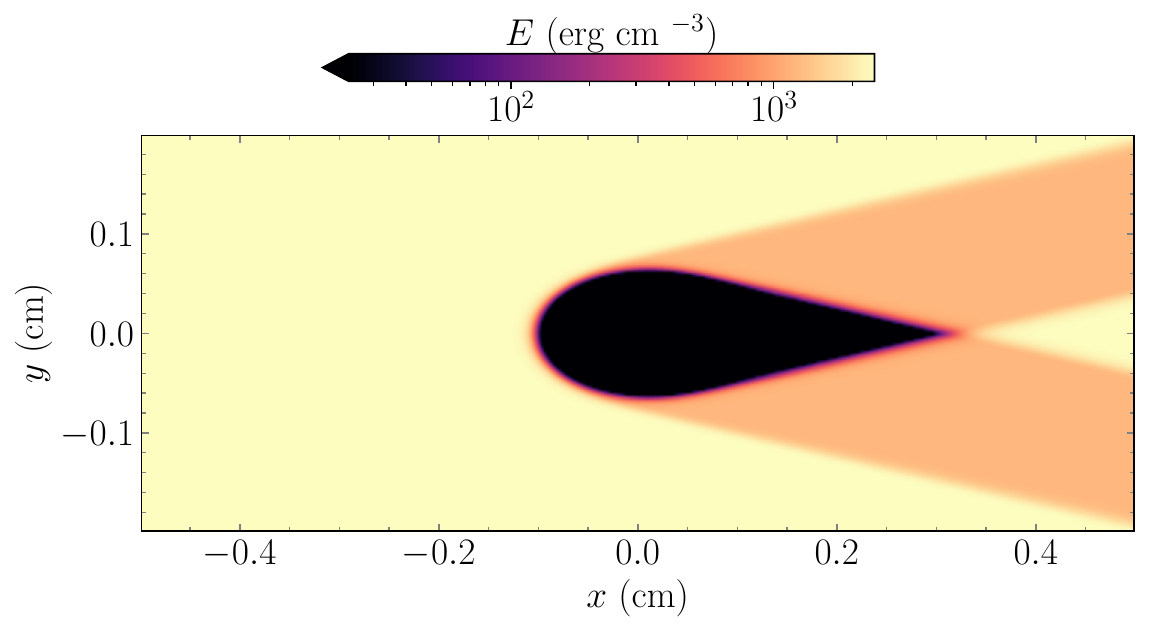}
\caption{Oblique shadows produced with the HM closure.}
\label{fig:shadowHM}
\end{figure}

We lastly showcase the HM method's ability to reproduce shadows cast by radiation emitted in different directions, either by separate sources or a single extended source. We considered the latter case by increasing the vertical domain to $[-0.2,0.2]$ with a resolution of $160 \times 160$, setting the left boundary fluxes as 
\begin{equation}
    \boldsymbol{F}_\pm = \frac{E_1}{2} \left(
    \sqrt{19/20},
    \pm\sqrt{1/20}
    \right)\,.
\end{equation}
In this way, radiation was simultaneously injected at the left boundary in two different directions. The resulting radiation energy distribution, shown in Fig. \ref{fig:shadowHM}, is the sum of the $E_+$ and $E_-$ distributions, each casting a shadow in its respective transport direction. As a result, the radiation field forms a shadow pattern with umbra, penumbra, and antumbra regions, characteristic of spatially extended light sources. These features cannot be reproduced with the M1 closure, which does not allow transport in multiple directions at the same location.

\section{Thermal modeling of protoplanetary disks}\label{S:Disks}

\begin{figure*}[t!]
\centering
\includegraphics[width=\linewidth]{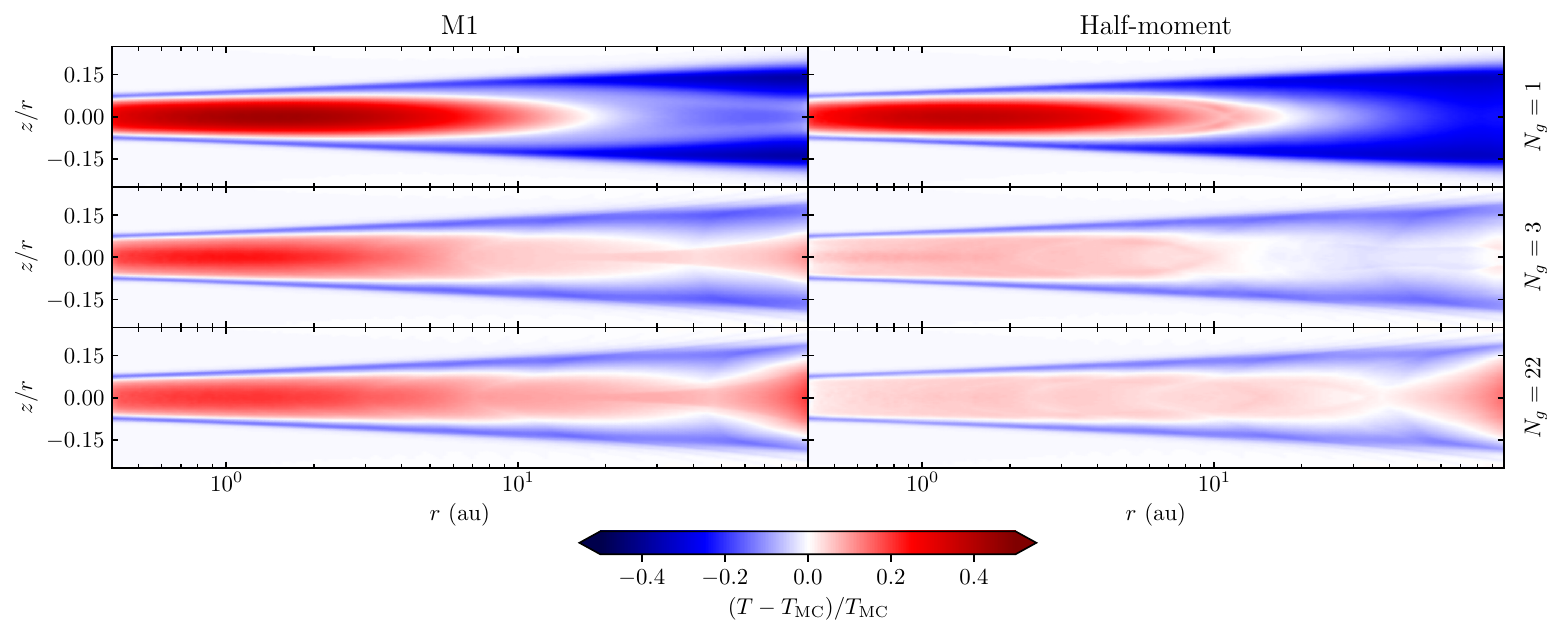}
\caption{Relative deviation of the disk temperatures from the MC distribution in the M1 and HM runs.}
\label{fig:Tcomparison2D}
\end{figure*}

We now investigate the accuracy of multigroup M1 and HM radiative transfer in an astrophysical scenario involving the computation of the temperature in a protoplanetary disk irradiated by a central T Tauri star. In particular, we evaluate whether our HM method can improve the agreement with MC models. We modeled the disk as a stationary axisymmetric gas and dust distribution with a fixed dust-to-gas mass ratio. In the M1 and HM runs, we used tabulated frequency-dependent dust opacities to compute both the stellar irradiation absorption term via ray tracing and Planck and Rosseland opacities for reprocessed radiation, using $N_g=1$, $3$, or $22$ frequency groups for the latter. At the same time, we adopted the same dust distribution, stellar parameters, and tabulated opacities to compute the temperature distribution with the MC radiative transfer code RADMC-3D \citep[][]{Dullemond2012}. Implementation details and simulation parameters for all methods are provided in Appendix \ref{A:DiskModel}.

\subsection{M1 simulations}\label{SSS:DiskM1}

\begin{figure*}[t!]
\centering
\includegraphics[width=\linewidth]{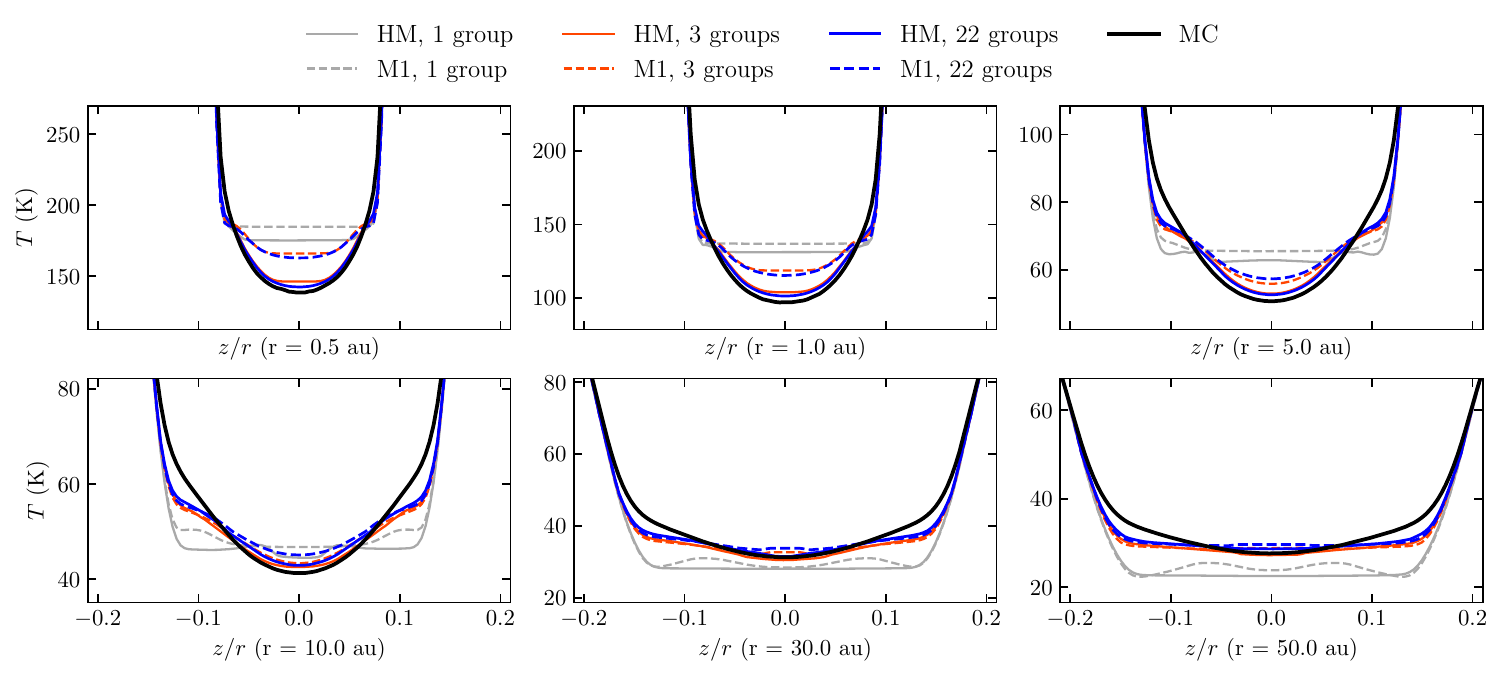}
\caption{Disk temperatures at fixed $r$ values in the MC, HM, and M1 runs.}
\label{fig:Tcomparison_slices}
\end{figure*}

In Figs. \ref{fig:Tcomparison2D} and \ref{fig:Tcomparison_slices}, we show a comparison of the temperature distributions obtained with the described methods in 2D maps and at fixed radii, respectively.  In the single-group M1 case, the midplane temperature is consistently overestimated in the optically thick region up to $44\%$ with respect to the MC distribution. There are two main reasons for this: the frequency averaging, addressed later in this section, and the interaction between incoming and outgoing photons. The latter phenomenon, discussed in Section 3.3 of \cite{MelonFuksman2022}, can be summarized as follows. Radiation is absorbed near the Planck-averaged radial $\tau=1$ surface for stellar photons (henceforth the irradiation surface) and re-emitted at a lower temperature both away and toward the disk. The flux emitted toward the disk encounters the flux escaping in the vertical direction, merging with it as exemplified in Section \ref{SS:FreeStreaming}. This reduces the flux leaving the disk, resulting in an overestimation of the midplane temperature. Additionally, this merger increases the radial flux produced by the radial temperature gradient in the optically thin region above the vertical $\tau=1$ surface. The resulting energy concentration in that region causes the deviation from vertical isothermality shown in Fig. \ref{fig:Tcomparison_slices} at large radii.

Above $10$ au, we obtained a significant underestimation of the disk temperature by up to $37\%$. This discrepancy is caused by an underestimation of the absorption opacity resulting from its Planck averaging. We verified this by computing the ratio between the single-group Planck opacity and its energy-averaged value for $N_g=22$, computed as $\kappa_E({N_g})=\sum_\ell \kappa_{P\ell} E_\ell/\sum_\ell E_\ell$. In the regions where the temperature is underestimated, we obtained $\kappa_P({N_g=1})/\kappa_E({N_g=22})<1$, with minimum values of $\sim1/5$. Consequently, for small $N_g$, the optically thin absorption of reprocessed radiation is underestimated, explaining why the temperature increases at large radii when increasing $N_g$.

In the single-group case, both the M1 and HM single-group runs produce vertically isothermal temperature distributions up to the irradiation surface. This is because, in stationary LTE, the flux source terms tend to zero, and therefore so does the divergence of the pressure tensor (Eqs. \eqref{Eq:HMRT_multigroup} and \eqref{Eq:M1RT}). Given that $\xi=1/3$ in this regime, the $E$ and $E_\pm$ gradients must also tend to zero. Since these energies are proportional to $T^4$ in LTE, the resulting temperature distribution is vertically isothermal.

Conversely, the multigroup runs result in U-shaped vertical temperature distributions in the optically thick regions \citep[see also][]{Pavlyuchenkov2025}. An explanation for this is provided in \cite{Dullemond2002}, and coincides with what we observe in our simulations. In these, the locations of the $\tau=1$ surfaces for vertically transported photons depend on opacity, resulting in departures from LTE at decreasing height for decreasing frequency. The disk cools via negative $G^0_{\ell}$ terms for the lowest-frequency groups, compensating for the escaping flux (Eq. \eqref{Eq:M1RT}). To maintain thermal balance, the sum of all $G^0_{\ell}$ must be zero below the irradiation surface (Eq. \eqref{Eq:M1RT_multigroup_MHD}), meaning that
the higher-frequency $G^0_{\ell}$ terms must increase. In other words, absorption must overcome emission at the higher-frequency groups in order to compensate for the energy loss at the low-frequency groups. This occurs through a steepening of the temperature gradient, which enhances the inward diffusive flux of the higher-frequency groups and consequently increases the inward-transported energy by exactly the amount needed to cancel the sum of all $G^0_{\ell}$.

The steepening of the temperature gradient in the multigroup runs results in lower midplane temperatures in optically thick regions compared to the single-group runs, leading to a better agreement with the MC distribution. With M1, we obtained maximum overestimations of the midplane temperature inside of $10$ au of $44\%$, $23\%$, and $21\%$ for $N_g=1$, $3$, and $22$, respectively. 

This agreement is aided by the fact that the frequency splitting prevents some of the interaction between photons entering and leaving the disk. Specifically, high-frequency inward-transported photons are emitted at the irradiation surface, while low-frequency outward-transported photons are emitted at the disk.
However, the temperature difference between the midplane and upper layers is not large enough to prevent middle-range frequency groups from containing both photons entering and leaving the disk (Fig. \ref{fig:groups}). Thus, regardless of the groups' definitions, frequency splitting alone is not sufficient to prevent the interaction between incoming and outgoing fluxes for all groups. As a result, the midplane temperature is still significantly overestimated in the optically thick regions.

\subsection{Half-moment simulations}\label{SSS:DiskHM}

In the HM case, the artificial interaction between outward and inward fluxes in the $\theta$ direction is absent, which improves the disk's cooling efficiency. This significantly improves the match with the MC distribution, leading to maximum temperature overestimations of $6\%$ and $8\%$ inside of $10$ au for $N_g=22$ and $3$, respectively. In the single-group case, the deviations are still large (up to $37\%$) since, as already discussed, the U shape cannot be recovered with a single frequency group. In conclusion, using $3$ frequency groups suffices to obtain similar errors as with a much finer frequency discretization (Fig. \ref{fig:Tcomparison2D}).

At large radii, the agreement between HM and MC is better for $N_g=3$ than for $N_g=22$. This results from the underestimation of the Planck opacity in that region for small $N_g$, which leads to a temperature decrease that compensates for the energy accumulation at $r=100$ au caused by boundary effects.

Despite the improved agreement with the MC distribution, the HM runs still overestimate the midplane temperatures in optically thick regions. Moreover, we also obtained a consistent temperature underestimation at the transition layers to the surface temperature. 

\begin{figure}[t!]
\centering
\includegraphics[width=\linewidth]{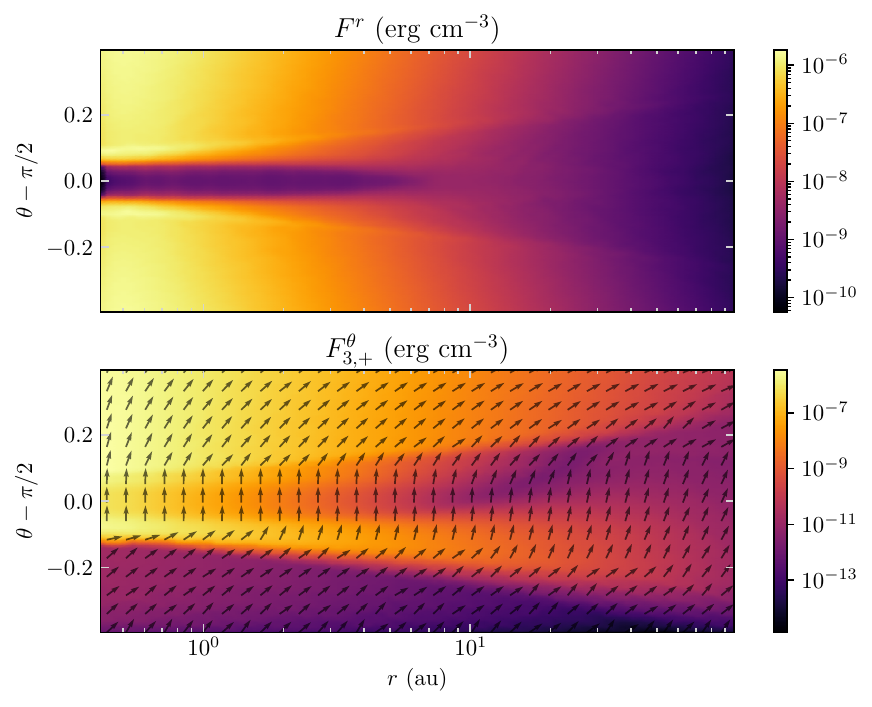}
\caption{
Fluxes in the HM disk simulation with $N_g=3$. Top: total radial flux. Bottom: $\theta$ component of the $+$ flux for the highest-frequency group ($\ell=3$),  also showing the direction of the $+$ flux for the same group.
}
\label{fig:Fr_F1plus}
\end{figure}

\begin{figure}[t!]
\centering
\includegraphics[width=\linewidth]{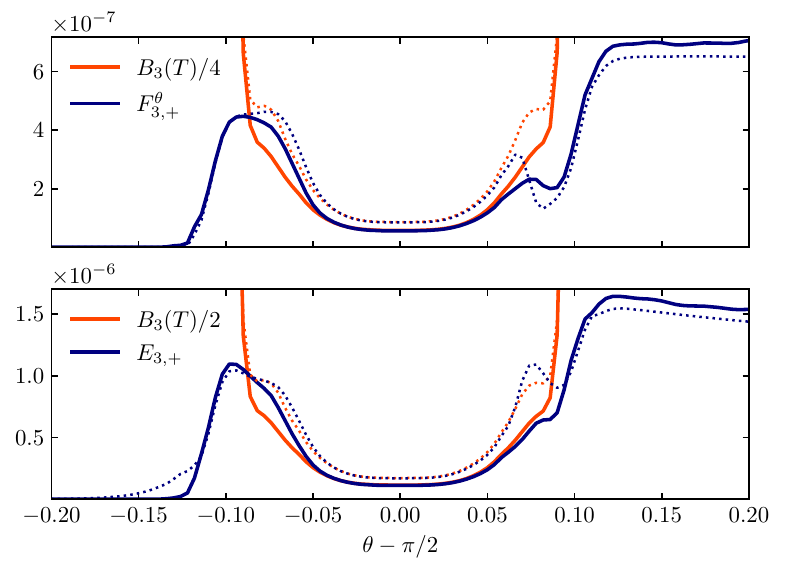}
\caption{
Half-moment ($+$) flux and energy density at $r=1$ au for the highest-energy group in the disk tests with $N_g=3$, given in erg cm$^{-3}$. Solid and dotted lines correspond to solutions obtained with and without enforcing Eq. \eqref{Eq:Fminv2}, respectively.
}
\label{fig:DiskFpEp}
\end{figure}

To understand these differences, we need to look at the flux distribution in the HM runs. In Fig. \ref{fig:Fr_F1plus}, we show the distribution of the total radial flux for $N_g=3$, together with the 
$F^{\theta}_{\ell,+}$ distribution for the highest-frequency group ($\ell=3$). Also shown is the direction of the vector field $\boldsymbol{F}_{3,+}$. This figure evidences that, as radiation escapes the disk, the vertical flux merges with the radial flux resulting from the radial temperature gradient. This results in a localized increase in the energy density above its LTE value below the irradiation surface, shown in Fig. \ref{fig:DiskFpEp} near $\theta-\pi/2\sim0.07$. As a result, according to Eq. \eqref{Eq:HMRT_multigroup}, the vertical flux gradient must become negative. This produces a significant decrease in the outward flux, shown in Fig. \ref{fig:DiskFpEp}, which reduces the cooling efficiency. This effect is further enhanced by the variation of the Eddington tensor as $f$ reaches small values significantly below $1/2$. In particular, the flux should in reality satisfy $f\geq1/2$ in that region, since $f$ should only increase from its LTE value of $1/2$ as energy is transported away from the disk in optically thin regions.

To minimize these effects, we enforced the physicality condition \eqref{Eq:physlimit2} via Eq. \eqref{Eq:Fminv2}. This slightly enhances the outward flux (see Fig. \ref{fig:DiskFpEp}), reducing the midplane temperature and improving the agreement with the MC distribution. For comparison, if Eq. \eqref{Eq:Fminv2} is not applied, we obtained maximum temperature overestimations within $10$ au of $44\%$, $17\%$, and $13\%$ for $N_g=1$, $3$, and $22$, respectively. Lastly, we also found that ignoring the constraint $D_{xz}\geq0$ (Eq. \eqref{Eq:Df_sm_1_2}) slightly increased the disk temperature, worsening the match with MC by a few percent.

\section{Discussion}\label{S:Discussion}

\subsection{Disk modeling}\label{SS:DiscussionDisk}

The multigroup HM temperatures are significantly closer to the MC distributions than those obtained with the M1 closure, even when using only three frequency groups. Compared to dynamical disk models relying on moment-based or single-group methods, this approach allows for more accurate modeling of HD and MHD disk processes self-consistently with radiative transfer without requiring high-resolution angular discretization. In particular, hydrodynamical instabilities are sensitive to temperature-dependent cooling timescales, and therefore the regions where they can operate have a strong dependence on the disk temperature \citep{Pfeil2019,MelonFuksman2024partII}. An accurate and self-consistent thermal modeling is also required for the precise modeling of disk observations \citep{Ziampras2023,Ziampras2024obs,Muley2024a}, planetary migration \citep{Guilera2021,Ziampras2024,Ziampras2025}, and shadowing effects \citep{MelonFuksman2022,Muley2024b}. 

While the application of Eq. \eqref{Eq:Fminv2} in our HM models minimizes the artificial reduction of the outgoing flux, it makes the method more diffusive in the $\hat{\boldsymbol{d}}$ direction, as characterized in Section \ref{SS:Shadow}. In the disk case, this prevents the formation of shadows cast by reprocessed radiation in the radial direction. It is important to note that this should not play a role in the modeling of phenomena produced by shadows cast by starlight \citep[e.g.,][]{MelonFuksman2022,Muley2024b}, as these are produced by the irradiation source term. Still, it remains to be explored if the interaction between radial and vertical fluxes can be reduced while also minimizing diffusivity.

In some cases, radiation-HD simulations with more accurate temperature distributions than in this work may be needed. In particular, not only the temperature but also its derivatives control the local growth rates of hydrodynamical instabilities, such as the vertical shear instability and the convective overstability \citep{Klahr2023a,Klahr2023b}. Naturally, one way to achieve this would be by means of frequency-dependent discrete ordinate methods. On the other hand, computing the temperature via MC methods provides an interesting alternative in scenarios where the temperature can be assumed constant or locally relaxed to an initial value. However, this approach is subject to the case-dependent drawbacks outlined in the introduction whenever compression and expansion work cannot be neglected. Alternatively, one could seek an approximate maximum-entropy closure for four quadrants, thus splitting the radial and vertical fluxes and removing their interaction. Additionally, frequency-dependent scattering, absent in this work to ease the method comparison, should also be included in realistic models. Further improvements could include obtaining separate different maximum-entropy closures for each frequency group (Section \ref{SS:MultigroupHM}), and including non-LTE emissivities \citep[e.g.,][]{Colombo2019}.

\subsection{Computational efficiency}\label{SS:DiscussionEfficiency}

Our method's computational efficiency is subject to its implementation. In particular, with an IMEX method, the overhead depends almost exclusively on the value of the reduced speed of light, which determines the time step for the radiation solver. Conversely, reducing the speed of light would not be necessary in a globally implicit implementation, at the cost of worse parallel scalability than IMEX methods.

For the disk test shown here with $\hat{c}/c= 2 \times 10^{-5}$, a typical minimum value for disk simulations with our chosen parameters, though problem-dependent, we found that solving the single-group radiation-HD problem advecting all three components of vector fields and using Newton's method in the implicit step is $\sim7$ times more expensive with M1 and 
$\sim 16$ times more expensive with HM than an equivalent HD simulation.
The cost depends on the convergence speed of the implicit step, so we measure it after reaching radiative equilibrium, representing a typical case for a radiation-HD simulation. 
When employing fixed-point solvers, these factors are slightly reduced to $6$ and $15$ for M1 and HM, respectively.
The difference between the Newton and fixed-point methods increases with $N_g$ as the cost of these methods for large $N_g$ is $\mathcal{O}(N_g^3)$ and $\mathcal{O}(N_g)$, respectively. For $N_g=22$ with HM, we obtained $t_\mathrm{HM}/t_\mathrm{HD}\sim795$ and $282$ with Newton and fixed point methods, respectively, while for M1 we obtained $t_\mathrm{M1}/t_\mathrm{HD}\sim120$ and $65$.

\begin{figure}[t!]
\centering
\includegraphics[width=\linewidth]{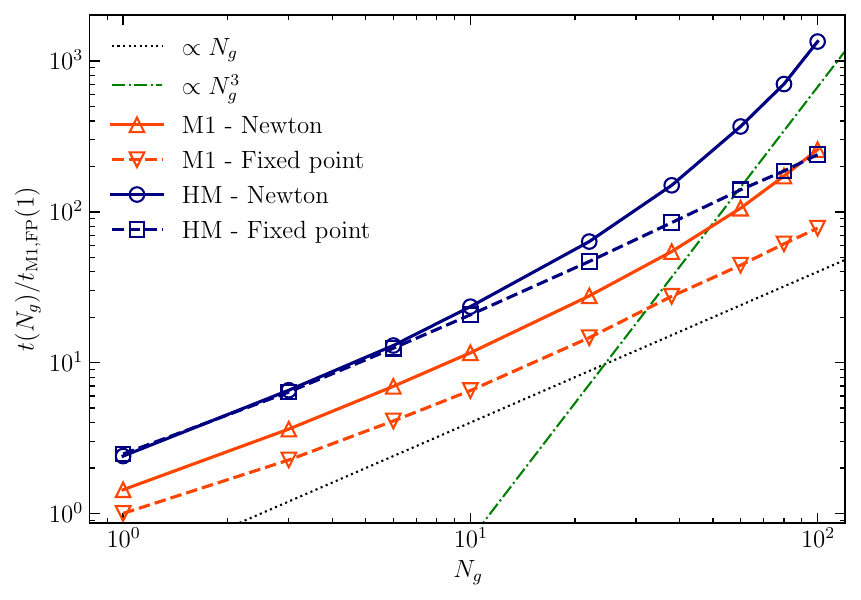}
\caption{
Computational cost of the radiation integration step using a single implicit iteration as a function of $N_g$, normalized to the fixed-point M1 cost for $N_g=1$.}
\label{fig:Ng_scaling}
\end{figure}

To characterize the scaling of both Newton and fixed-point methods for large $N_g$, we replicated the disk test with $N_g$ up to $100$ at a resolution of $100\times100$ using a single CPU (Fig. \ref{fig:Ng_scaling}). We computed the cost of the integration step fixing the number of iterations to $1$ in the implicit solver to avoid comparing runs with a different number of iterations. 
For low $N_g$, the cost of all methods is mostly due to the transport terms and therefore increases linearly with $N_g$, proportionally to the increase in the number of advected fields. The cost increase is slightly shallower for $N_g<3$ due to the overhead of operations involving the HD fields, despite the fact that we do not include HD advection. With Newton's method, the dependence on $N_g$ becomes steeper for larger $N_g$, eventually reaching $t\propto N_g^3$ when the overhead is dominated by the computational cost of the Gaussian elimination algorithm employed by the implicit solver. For the HM method, this occurs for lower $N_g$ than for M1 due to the higher cost of the implicit method, which iterates twice as many fields as in the M1 case. Since the average number of iterations in simulations is always $\geq 1$, Fig. \ref{fig:Ng_scaling} represents a best-case scaling scenario where the transition to the $\mathcal{O}(N_g^3)$ scaling occurs as late as possible.
On the contrary, the fixed point method remains linear for large $N_g$, leading to substantial speedup compared to Newton's method. In contrast, globally implicit methods \cite[e.g.,][]{Gonzalez2015} scale as $N_g^2$.

Our particular implementation via IMEX schemes can be compared with discrete ordinate methods such as that in \cite{Jiang2014}, which also relies on IMEX integration with a reduced speed of light. This method requires advecting as many fields as directions defined for the angle discretization. In that work, $8$ directions are sufficient to produce a slightly more diffusive crossing beam test compared to that shown in Fig. \ref{fig:crossing_beams}, obtained advecting $6$ fields  ($E_\pm,F^x_\pm$,$F^y_\pm$). While we have employed a larger resolution, it is likely that more directions are required in discrete ordinate methods to produce sharply transported light beams comparable to that figure, which would increase the overall cost compared to the HM method. In axisymmetric protoplanetary disk models with a single-frequency group using a similar globally implicit method by \cite{Jiang2021}, \cite{Zhang2024} found convergence with $16$ discrete directions, which is larger than the $6$ radiation fields advected in our single-group disk simulations. 
Generally, the required angular resolution depends on the problem at hand;
in particular, \cite{Zhu2020} used $80$ discrete directions in 3D models of FU Ori’s accretion disks. 

In general, discrete ordinate methods naturally outperform the HM closure's accuracy. For example, in the beam crossing test, they would allow for the addition of a third beam, which in the HM case would result in artificial interaction. Yet, our HM method can be preferable when desiring to prevent such interactions in a single chosen direction, or when requiring a high-resolution frequency discretization that renders a high-resolution angle discretization prohibitively expensive.

\section{Conclusions}\label{S:Conclusions}

We introduced an HM closure for radiative transfer that extends previous 1D HM formulations to multidimensional transport. In this strategy, a splitting direction is chosen, and half moments are defined by integrating the radiation specific intensity over each hemisphere defined by it. A closure for the half moments can then be obtained through maximum-entropy constraints. We derived our HM closure by finding an approximation to this maximum-entropy closure that agrees with it in the diffusion and free-streaming limits. The resulting closure is easy to implement, and the transport terms can be integrated using Riemann solvers with exactly computed eigenvalues.

We implemented this method as a modification of the M1 radiation-HD/MHD module in the open-source PLUTO code, relying on IMEX schemes to handle both the implicit integration of radiation-matter interaction terms and the explicit integration of transport terms. We also extended both the HM and M1 codes to support multiple frequency groups. We implemented Newton and fixed-point methods for the implicit step. For a large number of frequency groups $N_g$, the cost of these methods scales as $N_g^3$ and $N_g$, respectively.

In a series of numerical benchmarks, we demonstrated that the HM closure behaves as expected in both free-streaming and diffusive transport, including shadow-casting scenarios where radiation transitions between these regimes. In particular, the HM closure removes the artificial interaction between beams—unavoidable in the M1 closure—in selected spatial directions. This is achieved with minimal angular discretization, as integrating the specific intensity over two hemispheres suffices to produce sharp beams in arbitrary directions. Conversely, this artificial interaction cannot be avoided when the beams propagate in directions corresponding to the same half moment; that is, when their fluxes are simultaneously positive or negative along the splitting direction.

In models of protoplanetary disks around T-Tauri stars, the HM method outperforms the M1 closure by removing the artificial interaction between fluxes entering and leaving the disk. The significantly smaller temperature discrepancies compared to MC simulations represent an improvement over typical state-of-the-art moment-based and single-group dynamical simulations of circumstellar disks. In multigroup simulations with $22$ frequency groups, the HM closure limits the maximum temperature overestimation to $6\%$ compared to MC, whereas this overestimation increases to $21\%$ with the M1 closure. We found that as few as three groups are enough to reduce the temperature overestimation to $8\%$ with the HM closure. The remaining discrepancies with the MC distributions result from interactions between vertical and radial fluxes, which could be avoided using discrete ordinate methods or in future extensions to partial-moment closures over four quadrants.

\begin{acknowledgements}
We thank Anton Krieger and Shangjia Zhang for valuable comments and discussions that helped us improve this manuscript, as well as W. Bear for his support. M.F. acknowledges support from the European Research Council (ERC), under the European Union’s Horizon 2020
research and innovation program (grant agreement No. 757957). H.K.\ acknowledges funding from the European Research Council (ERC) via the ERC Advanced Grant “TiPPi - Turbulence, Pebbles and Planetesimals: The Origin of Minor Bodies in the Solar System” (project ID 855130) and from the Heidelberg Cluster of Excellence (EXC 2181 - 390900948) “STRUCTURES: A unifying approach to emergent phenomena in the physical world, mathematics, and complex data,” funded by the German Excellence Strategy. Numerical simulations were run on MPIA's VERA cluster, hosted at the Max-Planck Computing and Data Facility in Garching, Germany. Data analysis and plotting were performed using the Numpy \citep{Harris2020}, Matplotlib \citep{Hunter2007}, and PyPLUTO \citep{Mattia2025pyPLUTO} Python libraries. 
\end{acknowledgements}

\bibliographystyle{aa}
\bibliography{refs}

\begin{appendix}
\section{Comparison of the approximate and exact half-moment closures}\label{A:approximate_exact}

\begin{figure*}[t]
\centering
\includegraphics[width=\linewidth]{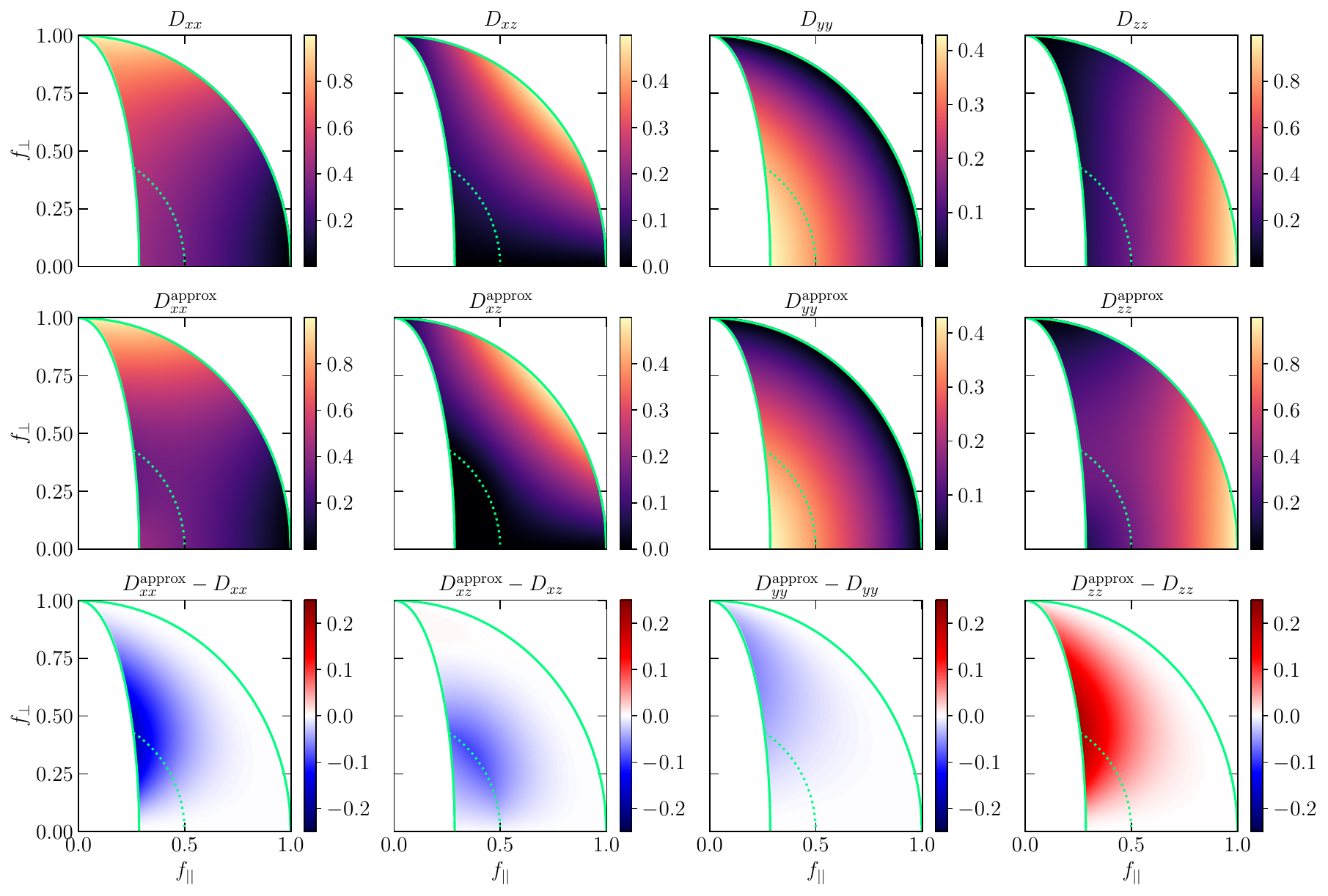}
\caption{
Nonzero components of the Eddington tensor in the exact maximum-entropy HM closure ($D_{ij}$) and the approximate closure introduced in this work ($D^\mathrm{approx}_{ij}$). The bottom row shows the difference between the two. Solid lines highlight the bounds given by Eqs. \eqref{Eq:physlimit1} and \eqref{Eq:physlimit2}, while dotted lines correspond to $f=1/2$. The isotropic and free-streaming limits correspond to $(f_{||},f_\perp)=(1/2,0)$ and $f=1$, respectively. Here $(f_{||},f_\perp)$ is defined as $(f_z,f_x)$.
The components shown are computed assuming zero $y$-flux, which suffices to represent all other flux directions, as is shown in Appendix \ref{A:approximate_exact}.
}
\label{fig:closure}
\end{figure*}

Here we assess the accuracy of the HM closure introduced in this work compared with the exact maximum-entropy HM closure resulting from a numerical inversion of the Lagrange multipliers (Section \ref{SS:maxentrHM}). To do this, we need to compare the Eddington tensor $\mathbb{D}(E,\boldsymbol{F})$ obtained in both approaches. Here, as in the rest of this appendix, we omit the subscripts $\pm$ to simplify the notation.

This comparison can be simplified by noting that both the exact and approximate closures satisfy the same transformation law for $\mathbb{D}$ when $\boldsymbol{F}$ is rotated about $\hat{\boldsymbol{d}}$ according to a rotation matrix, $R_{\hat{\boldsymbol{d}}}$; namely,
\begin{equation}
\mathbb{D}(E,R_{\hat{\boldsymbol{d}}}\,\boldsymbol{F})
=
R_{\hat{\boldsymbol{d}}}\,
\mathbb{D}(E,\boldsymbol{F})\,
R_{\hat{\boldsymbol{d}}}^\intercal\,.
\end{equation}
Thus, a comparison for $\boldsymbol{F}$ in any plane containing $\hat{\boldsymbol{d}}$ suffices to compare these closures for arbitrary $(E,\boldsymbol{F})$. 
One way to prove this is to start from the computation of the half moments from Eq. \eqref{Eq:LagrangeHM} to show that the transformation $\boldsymbol{b}\rightarrow R_{\hat{\boldsymbol{d}}} \boldsymbol{b}$ results in the transformations
\begin{equation}\label{Eq:RotationHM}
    E \rightarrow E
    \,,\,\,\,\,\,
    \boldsymbol{F} \rightarrow R_{\hat{\boldsymbol{d}}}\boldsymbol{F}
    \,,\,\,\,\,\,
    \mathbb{D} \rightarrow
    R_{\hat{\boldsymbol{d}}}\,
    \mathbb{D}\,
    R_{\hat{\boldsymbol{d}}}^\intercal\,.
\end{equation}
This means that the transformation of $\mathbb{D}$ as a function of the Lagrange multipliers when $\boldsymbol{b}\rightarrow R_{\hat{\boldsymbol{d}}}\boldsymbol{b}$ is equivalent to its transformation as a function of $\boldsymbol{F}$ when $\boldsymbol{F}\rightarrow R_{\hat{\boldsymbol{d}}}\boldsymbol{F}$. An alternative way is to view $R_{\hat{\boldsymbol{d}}}$ as a transformation of coordinates, noting that the rotation symmetry about $\boldsymbol{\hat{d}}$ of the radiation fields is not broken by the choice of the splitting direction.

To perform the comparison, we define a basis $\{\hat{\boldsymbol{x}},\hat{\boldsymbol{y}},\hat{\boldsymbol{z}}\}$ of $\mathbb{R}^3$ such that, without loss of generality, $\hat{\boldsymbol{d}}=\hat{\boldsymbol{z}}$ and $\boldsymbol{F}$ is in the $xz$ plane. Therefore, $\boldsymbol{b}$ must also be contained in that plane\footnote{
Using spherical coordinates, it is straightforward to prove that $b_y=0$ implies $F_y=0$. Since $\boldsymbol{F}$ and $\boldsymbol{b}$ transform identically under $R_{\hat{\boldsymbol{d}}}$, they must always lie in the same plane containing $\hat{\boldsymbol{d}}$.
}, and so we can parameterize it as $\boldsymbol{b}=b\,(\sin \beta, 0, \cos \beta)$, with $0\leq b < 1$ and $0\leq \beta \leq \pi$. This parameter reduction largely simplifies the calculation of the $\mathbb{D}$ components in the exact maximum-entropy closure, which can be obtained as $D^{ij}=P^{ij}/E$ and displayed as functions of $f_{||}=F_z/E$ and $f_\perp=F_x/E$, computing all of these quantities as functions of $(b,\beta)$.

While the $\mathbb{D}$ components are always nonnegative in this basis, $D^\mathrm{approx}_{xz}$ defined by Eqs. \eqref{Eq:EddTensor} and \eqref{Eq:xiHM} is negative for ${f<1/2}$, with minimum values close to $-0.1$.
To improve the match between the approximate and exact closures, we enforce $D^\mathrm{approx}_{xz}\geq0$ in the defined basis by zeroing ${(3\xi-1)/2}$ in the diagonal terms for $f<1/2$.

The resulting Eddington tensor for $f<1/2$ can be expressed in coordinate-independent form as
\begin{equation}\label{Eq:Ddiag}
    \mathbb{D}^\mathrm{approx}=
    \frac{1-\xi}{2}\mathbb{I}
    +\frac{3\xi-1}{2}
    \left[
    (\hat{\boldsymbol{n}}\cdot\hat{\boldsymbol{x}})^2
    \hat{\boldsymbol{x}} \hat{\boldsymbol{x}}^\intercal
    +
    (\hat{\boldsymbol{n}}\cdot\hat{\boldsymbol{d}})^2
    \hat{\boldsymbol{d}} \hat{\boldsymbol{d}}^\intercal
    \right]\,,
\end{equation}
where $\mathbb{I}$ is the identity tensor and we have already zeroed $D_{xz}$. Since $\hat{\boldsymbol{x}}$, $\hat{\boldsymbol{n}}$, and $\hat{\boldsymbol{d}}$ are coplanar, we have $\hat{\boldsymbol{x}}=k_1 \hat{\boldsymbol{n}} + k_2 \hat{\boldsymbol{d}}$. Together with the orthonormality conditions $\hat{\boldsymbol{x}}\cdot \hat{\boldsymbol{d}}=0$ and $\hat{\boldsymbol{x}}\cdot \hat{\boldsymbol{x}}=1$, this results in ${\hat{\boldsymbol{x}}={\pm(\hat{\boldsymbol{n}}-(\hat{\boldsymbol{n}}\cdot \hat{\boldsymbol{d}})\hat{\boldsymbol{d}})}/(1-(\hat{\boldsymbol{n}}\cdot \hat{\boldsymbol{d}})^2)^{1/2}}$\,. Lastly, replacing this expression in Eq. \eqref{Eq:Ddiag}, we obtain the general form of $\mathbb{D}^\mathrm{approx}$ for $f<1/2$ (Eq. \eqref{Eq:Df_sm_1_2}).

In Fig. \ref{fig:closure}, we show the nonzero components of $\mathbb{D}$ and $\mathbb{D}^\mathrm{approx}$, along with their differences, as functions of $(f_{||},f_\perp)$. 
We obtain that both closures have similar functional forms, coinciding near the free-streaming and diffusion limits.
Away from these limits, differences are typically under $10 \%$ except in some regions near the limit given by Eq. \eqref{Eq:physlimit2}. In particular, $D_{zz}$ is overestimated up to a factor of $\sim 2$ in a narrow band near that limit, and $D_{xz}$ is underestimated by a similar factor away from the isotropic limit when $f\sim1/2$.

Often the radiation field is either near the diffusion or the free-streaming regime (Fig. \ref{fig:shadow_histograms}), where the approximate and exact maximum-entropy closures coincide. However, in some cases, it can reach $(f_{||},f_\perp)$ regions where differences are maximal, particularly in scenarios involving beam crossing (Section \ref{SS:FreeStreaming}). In such pathological cases, inaccuracies stem from the HM approach itself, regardless of whether the exact or approximate closures are used. Therefore, both closures should fail in similar scenarios and perform similarly whenever such effects are absent.

\section{Radiation-matter interaction terms}\label{A:sourceterms}

For each frequency, monochromatic HM radiative transfer equations can be obtained by multiplying Eq. \eqref{Eq:RTE} by $(1,\hat{\boldsymbol{n}})$ and integrating over $\mathcal{S}_\pm$. This leads to the following system of equations:
\begin{equation}
    \begin{split}
        \frac{1}{\hat{c}}\partial_t E_{\nu,\pm} + \nabla\cdot \boldsymbol{F}_{\nu,\pm} &=
        \rho\kappa_\nu\left(
        2 \pi B_\nu(T) - E_{\nu,\pm}        
        \right) \\
        & + \rho\sigma_\nu\left(
        \frac{E_\nu}{2} - E_{\nu,\pm}        
        \right)
        \\
        \frac{1}{\hat{c}}\partial_t \boldsymbol{F}_{\nu,\pm} + \nabla\cdot \mathbb{P}_{\nu,\pm} &=
        - \rho \chi_\nu \boldsymbol{F}_{\nu,\pm} \\
        & \pm \frac{\rho}{4} \left(
        \kappa_\nu\, 4 \pi B_\nu(T) 
        + \sigma_\nu E_\nu 
        \right) \hat{\boldsymbol{d}}
        \,,
    \end{split}
\end{equation}
where $E_{\nu,\pm}$, $\boldsymbol{F}_{\nu,\pm}$, and $\mathbb{P}_{\nu,\pm}$ are monochromatic half moments and $E_\nu=E_{\nu,+}+E_{\nu,-}$. Here we have replaced $c\rightarrow \hat{c}$ in the factors multiplying the time derivatives. Opacity coefficients are computed in the fluid's comoving frame, and relativistic corrections are ignored.

The frequency integration of the first of these equations is straightforward if we replace $\kappa_\nu$ and $\sigma_\nu$ with their Planck-averaged opacities, with the criterion that this is a valid approximation when $I_\nu \sim B_\nu(T)$ \citep{Mihalas}. This results in the $G^0_{\ell,\pm} $ term in Eq. \eqref{Eq:HMSourceTerms_grey_1}.

On the other hand, to average the second equation in such a way that the flux is accurate in the diffusion regime, we need to use Rosseland-averaged opacities. This is simple in the pure-absorption and pure-scattering cases, where it suffices to replace $\kappa_\nu\rightarrow \kappa_{R\ell}$ or $\sigma_\nu\rightarrow \sigma_{R\ell}$, respectively. This results in the source terms in Eqs. \eqref{Eq:GpmAbsorption} and \eqref{Eq:GpmScattering}. For combined absorption and scattering, we can follow similar steps to the usual derivation of the Rosseland opacity \citep{Mihalas} by zeroing the time derivatives and assuming $I_\nu\sim B_\nu(T)$ to compute the pressure tensor in the diffusion regime, resulting in $\nabla \cdot \mathbb{P}_{\nu,\pm} \approx \frac{2 \pi}{3}\nabla  B_\nu(T)$. In this way, we obtain the frequency-integrated source term
\begin{align}
    \boldsymbol{G_{\ell,\pm} } &=
    \rho \chi_{R\ell}
    \left(
    \boldsymbol{F}_{\ell,\pm} 
    \mp \frac{B_\ell(T)}{4}\hat{\boldsymbol{d}}
    \right) \nonumber \\
    &\pm
    \frac{\rho \chi_{R\ell}}{4 c}
\int^{\nu^\mathrm{max}_{\ell}}_{\nu^\mathrm{min}_{\ell}}\mathrm{d}\nu 
    \,\frac{\sigma_\nu}{\chi_\nu}
    \left(
    4 \pi B_\nu(T)-
    E_\nu
    \right) \label{eq:Gpm_2} \\
    &=
    \rho \chi_{R\ell}
    \left(
    \boldsymbol{F}_{\ell,\pm} 
    \mp \frac{E_\ell}{4}\hat{\boldsymbol{d}}
    \right) \nonumber \\
    &\pm
    \frac{\rho \chi_{R\ell}}{4 c}
  \int^{\nu^\mathrm{max}_{\ell}}_{\nu^\mathrm{min}_{\ell}}\mathrm{d}\nu 
    \,\frac{\kappa_\nu}{\chi_\nu}
    \left(
    E_\nu-
    4 \pi B_\nu(T)
    \right)\,. \label{eq:Gpm_4}
\end{align}
Since $E_\nu\sim 4\pi B_\nu(T)$ in the diffusion regime and $\chi_{R\ell}$ is small away from it, the integrals on the right-hand side can likely be ignored, leaving only the first term. Under this criterion, it should be convenient to use Eq. \eqref{eq:Gpm_2} for absorption-dominated problems and Eq. \eqref{eq:Gpm_4} for scattering-dominated ones. Alternatively, one could either Planck-average $\sigma_\nu/\chi_\nu$ in Eq. \eqref{eq:Gpm_2} or  $\kappa_\nu/\chi_\nu$ in Eq. \eqref{eq:Gpm_4}. While this is not necessary in this work as we only use frequency-dependent opacities for absorption-only problems (Section \ref{S:Disks}), future studies should compare these approaches in problems involving both frequency-dependent absorption and scattering.\\
\section{Signal speeds}\label{A:SignalSpeeds}

Signal speeds are computed as the eigenvalues of the Jacobian matrices of Eq. \eqref{Eq:hyperbolicHM}. This calculation can be significantly simplified by writing the system in Cartesian coordinates and considering transport along single directions. Taking $x$ as the chosen direction without loss of generality, we consider the following system:
\begin{equation}
    \frac{1}{\hat{c}}\partial_t \mathcal{U} + \partial_x \mathcal{F}^x = 0\,,
\end{equation}
which has the Jacobian $J^x_{ij}={\partial \mathcal{F}^x_i  / \partial \mathcal{U}^j }$, where $\mathcal{F}^x=(F^x,P^{xx},P^{xy},P^{xz})$. Here, and in the rest of this section, we omit the group and sign subindices $(\ell,s)$. 

Here we show the general result of this calculation for transport in an arbitrary direction $\hat{\boldsymbol{e}}_i$ aligned with a coordinate axis. 
As in the M1 closure, the resulting eigenvalues for $f\geq1/2$ only depend on $f$ and $\mu = \hat{\boldsymbol{e}}_i \cdot \hat{\boldsymbol{n}}$. We obtain a set of three eigenvalues, $\{\lambda_1,\lambda_2,\lambda_3\}$, where $\lambda_2$ is twice degenerate. In units of $\hat{c}$, these are of the form
\begin{equation}
    \begin{split}
        \lambda_1 &= \beta_1 \mu - \sqrt{\beta_0+\beta_2 \mu^2} \\
        \lambda_2 &= \frac{6f-1-\sqrt{\alpha}}{4 f} \mu \\
        \lambda_3 &= \beta_1 \mu + \sqrt{\beta_0+\beta_2 \mu^2} \\
    \end{split}\,,
\end{equation}
where
\begin{equation}
    \begin{split}
    \alpha(f) &= 1 + 12 f (1-f)\\
        \beta_0(f) &= \frac{1-\xi}{2} + \left( f + \frac{1-3 \xi}{2 f} \right) \frac{\xi'}{2} \\
        \beta_1(f) &= \frac{\xi'}{2} \\
        \beta_2(f) &= \frac{(\xi')^2}{4} - f \xi' + \xi - \beta_0 \,.
    \end{split}
\end{equation}
Here $\xi$ is defined in Eq. \eqref{Eq:xiHM}, while $\xi'$ denotes its derivative,
\begin{equation}
    \xi'(f) = \frac{d \xi}{d f} = \frac{1 - \sqrt{\alpha} + 2 f (6-6f + \sqrt{\alpha})}{\alpha}\,.
\end{equation}

When $D_{xz}=0$ is enforced for $f<1/2$, the rotation symmetry of $\mathbb{D}$ in arbitrary directions is lost, leading to different eigenvalues for $\hat{\boldsymbol{e}}_i$ parallel and orthogonal to $\hat{\boldsymbol{d}}$. In both cases, the eigenvalues are not necessarily bound by $\pm\hat{c}$, and are complex for some values of $f$ and $\hat{\boldsymbol{n}}$, meaning that the system is not always strictly hyperbolic. Instead of computing the eigenvalues exactly, we maintain the presented form of the signal speeds also for $f<1/2$, with the advantage that the resulting speeds are continuous at $f=0.5$. We also enforce $|\lambda_i|\leq\hat{c}$ to prevent superluminal transport for some $\mu$ values when $f\rightarrow2/7$. Improvements to these estimates would require an exhaustive analysis of the system's characteristic waves for $f<1/2$, which we defer to future investigations.

In highly optically thick regions, signal speeds must be reduced to prevent excessive numerical diffusion. To do this, we adopt the prescription in Eq. 47 of \cite{Sadowski2013}.

\begin{figure}[t]
\centering
\includegraphics[width=\linewidth]{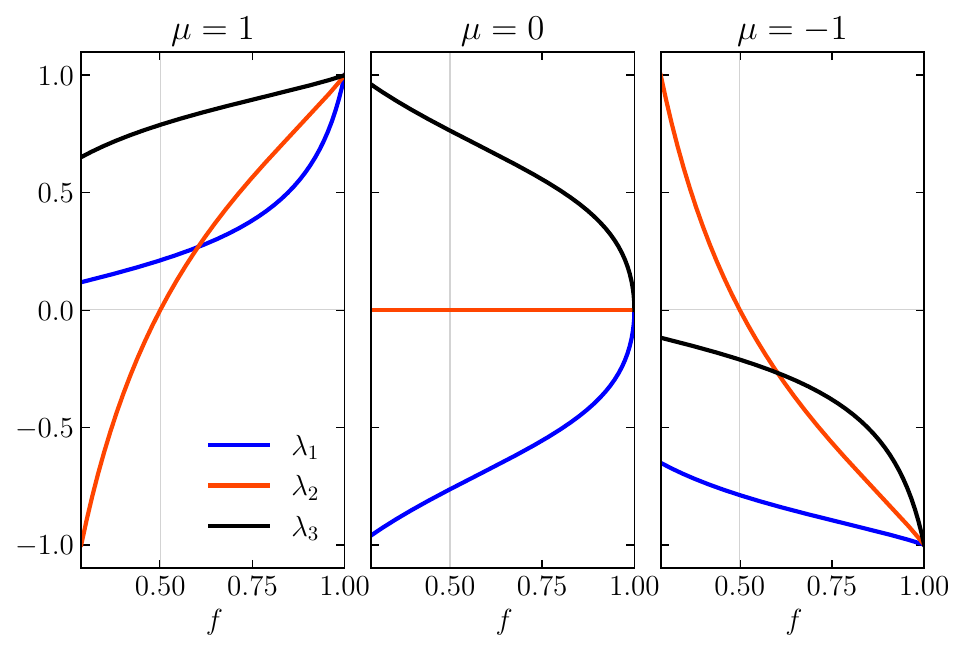}
\caption{
Eigenvalues of the HM closure for $\boldsymbol{F}_\pm$ aligned parallel ($\mu=1$), perpendicular ($\mu=0$), and antiparallel ($\mu=-1$) to the propagation direction.
}
\label{fig:eigenvalues}
\end{figure}

\section{Implicit integration of source terms}\label{A:ImplicitMethod}

The implicit methods presented here are analogous to those used in the PLUTO radiation module \citep[see][Appendix B]{MelonFuksman2022}. The energy evolution equations (Eq. \eqref{Eq:HMRT_implicit}), together with the conservation of $E_\mathrm{tot}$ (Eq. \eqref{Eq:Etot_mtot}), are used to write one of the energies, either $E_{\ell,s}$ or $\mathcal{E}$, as a function of the remaining ones, yielding a system of $2 N_g$ equations with $2 N_g$ unknowns. In our implementation, we can either choose $V_1=\{E_{\ell,s}\}$ as the iterated variables in Newton's method, or $V_2=\{\epsilon,E_{\ell,s\neq \ell_0,s_0}\}$ for a chosen $(\ell_0,s_0)$, where $\epsilon = (\mathcal{E}-\rho \boldsymbol{v}^2/2)/\rho$ is the specific internal energy. To ensure convergence, it is convenient to exclude the largest energy density from the iterated variables. This helps prevent small relative errors in a large variable from becoming large relative errors in smaller variables, which would require additional iterations to reach convergence.

We discretize Eq. \eqref{Eq:HMRT_implicit} implicitly by evaluating source terms at the next time step as
\begin{equation}
    E_{\ell,s}  = E_{\ell,s}' - \Delta t \hat{c}\, G^0_{\ell,s}\,,
\end{equation}
where we use tildes to denote fields evaluated at the previous time step, while untilded terms are evaluated at the next time step. Expanding this expression, we obtain
\begin{equation}\label{Eq:ImplEqEnergy}
    0 = E'_{\ell,s} + \zeta^\kappa_\ell \frac{B_\ell(T)}{2}
    - \left(1+\zeta^\kappa_\ell+\frac{\zeta^\sigma_\ell}{2}\right) E_{\ell,s}
    + \frac{\zeta^\sigma_\ell}{2} E_{\ell,-s}
    \,,
\end{equation}
where we have defined $\zeta^\kappa_\ell =\Delta t \hat{c} \rho \kappa_\ell$ and $\zeta^\sigma_\ell =\Delta t \hat{c} \rho \sigma_\ell$. If the set of variables $V_1$ is iterated, $T(\epsilon)$ is obtained as a function of $V_1$ using energy conservation ($E_\mathrm{tot}=E_\mathrm{tot}'$ and Eqs. \eqref{Eq:Etot_mtot}). Conversely, if the variables $V_2$ are iterated, we use energy conservation to obtain $E_{\ell_0,s_0}$ as a function of $V_2$.

The resulting system of equations is solved via a standard Newton method. 
Variable opacities, depending for instance on the temperature, are updated at each iteration and treated as constants for the computation of the Jacobian coefficients, in such a way that these have a simple polynomial form. To improve the scaling with $N_g$, we also implemented an alternative fixed-point algorithm by using Eq. \eqref{Eq:ImplEqEnergy} to express each $E_{\ell,s}$ as a function of $T$. The algorithm iterates all $E_{\ell,s}$ using these expressions and updates $T$ via energy conservation.

After each iteration, fluxes are updated as
\begin{equation}\label{Eq:implFls}
    \boldsymbol{F}_{\ell,\pm} = 
    \frac{1}{1+\zeta^\chi_\ell}\left[
    \boldsymbol{F}'_{\ell,\pm} 
    \pm
    \zeta^\kappa_\ell\frac{B_\ell(T)}{4}
    \hat{\boldsymbol{d}}
    \pm
    \zeta^\sigma_\ell \frac{E_\ell}{4}\hat{\boldsymbol{d}}
    \right]\,,
\end{equation}
with $\zeta^\chi_\ell = \Delta t \hat{c} \rho \chi_\ell$, while momentum is updated in such a way that the modified total momentum defined in Eq. \eqref{Eq:Etot_mtot} is conserved. If opacities are computed via frequency-averaging within each frequency group, the $\zeta$ functions in this expression use the Rosseland-averaged values of $\kappa$, $\sigma$, and $\chi$. In that case, 
Eq. \eqref{Eq:implFls} is modified according to the form of the flux source terms in the pure-absorption, pure-scattering, and combined absorption and scattering cases (Appendix \ref{A:sourceterms}). Conversely, Eq. \eqref{Eq:ImplEqEnergy} uses the Planck-averaged values of $\kappa$ and $\sigma$.

If external irradiation is included, the corresponding source term is simply integrated by adding ${-\Delta t \nabla\cdot\boldsymbol{F}_\mathrm{Irr}}$ to $E_\mathrm{tot}$, as done in \cite{Muley2023}. Compared to adding this term explicitly during the HD/MHD integration, this choice ensures faster convergence, as the initial guess for the iterated fields is closer to their converged values.

\section{Additional tests}\label{A:AdditionalTests}

\subsection{Crossing fluxes in a purely absorbing medium}\label{SS:1Dcrossing}

To test the joint implementation of transport and interaction terms, we reproduced the crossing fluxes test in \cite{Frank2006}. We considered a purely absorbing medium by switching off the update of the gas temperature. In the 1D domain $[-L/2,L/2]=[-0.5,0.5]$, discretized with $1000$ cells, we initialized the energy density to $10^{-7}$ in LTE. The density and absorption opacity are fixed to $\rho=1$ and $\kappa=2.5$. At $t=0$, freely streaming radiation is injected at both domain boundaries with $E=E_1=1.1357\times 10^4$, chosen to emulate the parameters in the cited work.

This problem is equivalent to considering a frequency-integrated radiative intensity $I=c E_1 \delta(\hat{\boldsymbol{n}}\pm\hat{\boldsymbol{x}})$ at $x=\pm L/2$. Fixing these values at the boundaries, we can solve the stationary frequency-integrated RTE to obtain
${I(x,\hat{\boldsymbol{n}}) = c E_1 \left(
    \delta(\hat{\boldsymbol{n}}-\hat{\boldsymbol{x}})
    e^{-\rho \kappa (x+L/2)}
    +
    \delta(\hat{\boldsymbol{n}}+\hat{\boldsymbol{x}})
    e^{-\rho \kappa (x-L/2)}
    \right)}$,
which translates into a radiation energy density of the form
\begin{equation}\label{Eq:crossingfluxes_exact}
    E(x) = E_1 \left(
    e^{-\rho \kappa (x+L/2)}
    + e^{-\rho \kappa (x-L/2)}
    \right)\,.
\end{equation}
This also coincides with the exact solution of the stationary HM equations (Eq. \eqref{Eq:HMRT_multigroup}) in the free-streaming limit ($f=1$). In contrast, the M1 closure deviates from this functional form, as it cannot simultaneously handle leftward and rightward transport at the same location.

\begin{figure}[t!]
\centering
\includegraphics[width=\linewidth]{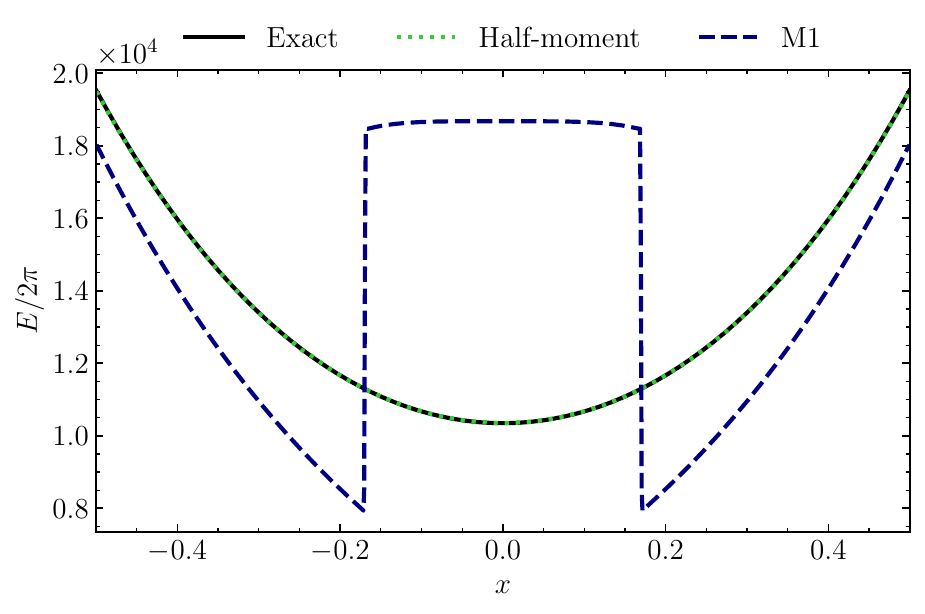}
\caption{Stationary energy density distribution in the crossing fluxes test in a purely absorbing medium. The solutions produced with the HM and M1 closures, normalized by $2\pi$ for comparison with \cite{Frank2006}, are compared with the exact solution of the RTE (Eq. \eqref{Eq:crossingfluxes_exact}).}
\label{fig:1Dcrossing}
\end{figure}

We show these differences in Fig. \ref{fig:1Dcrossing}, which compares the exact solution with the stationary radiation energy distributions obtained with our HM scheme (taking $\hat{\boldsymbol{d}}=\hat{\boldsymbol{x}}$) and the M1 closure. While the HM distribution matches the exact solution, the M1 closure merges the converging beams, resulting in an artificial energy accumulation at the domain center and a departure from the exact solution in the entire domain.

\subsection{Diffusion}\label{SS:Diffusion}

\subsubsection{Pure scattering}\label{SS:Diffusion_absorption}

In this section, we investigate diffusive transport in a constant opacity 1D medium, beginning with the pure-scattering radiative pulse test in \cite{Sadowski2013}. We consider the domain $[-100,100]$ cm discretized with $201$ cells, where we set constant $\rho\chi=\rho\sigma=$ 10 cm$^{-1}$ and $\kappa = 0$. We switch off the evolution of matter fields and set the initial radiation energy density as $E=a_R T^4$, with an effective temperature defined as ${T=T_0 \left(1 + A e^{-(x/w)^2}\right)}$, where $A=100$, $T_0=10^4$ K, and $w=15$ cm. Choosing $\hat{\boldsymbol{d}}=\hat{\boldsymbol{x}}$, the radiation fields are initialized in LTE, with $E_\pm = E/2$ and $\boldsymbol{F}_\pm = \pm (E/4) \hat{\boldsymbol{d}}$.

The optical depth across the Gaussian width at half maximum is $\sim 250$, which is sufficiently high for the radiation field to evolve in the diffusion regime with $\xi\approx 1/3$. This means that the equilibration time between the interaction and transport terms is much shorter than the timescale over which the radiation fields change, 
which allows us to drop the time derivatives in the evolution equations for $\boldsymbol{F}_\pm$. This yields the following equation for the diffuse fluxes:
\begin{equation}\label{Eq:diffuseFpm}
    \boldsymbol{F}_\pm = \pm \frac{E}{4} \hat{\boldsymbol{x}}
    -\frac{1}{3 \rho \chi} \nabla E_\pm\,.
\end{equation}
Lastly, summing the $+$ and $-$ evolution equations for the fluxes and energy densities (Eq. \eqref{Eq:HMRT_multigroup}), the diffusion equation for the total energy is recovered:
\begin{equation}\label{Eq:diffuseE}
    \partial_t E = 
    \frac{\hat{c}}{3 \rho \chi} \nabla^2 E\,.
\end{equation}
This equation can be solved by means of a Fourier transformation in the spatial domain, resulting in the following exact solution:
\begin{equation}
    E(t,x) = a_R T_0^4 \left[
    1 + 
    \sum^{4}_{n=1} \binom{4}{n} A^n
    \sqrt{\varphi(n,t)}
    e^{-n \varphi(n,t) (x / w)^2 }
    \right]\,,
\end{equation}
where
\begin{equation}
    \varphi(n,t) = \frac{3 \chi w^2}{3 \chi w^2 + 4 \hat{c} n t}\,.
\end{equation}

\begin{figure}[t!]
\centering
\includegraphics[width=\linewidth]{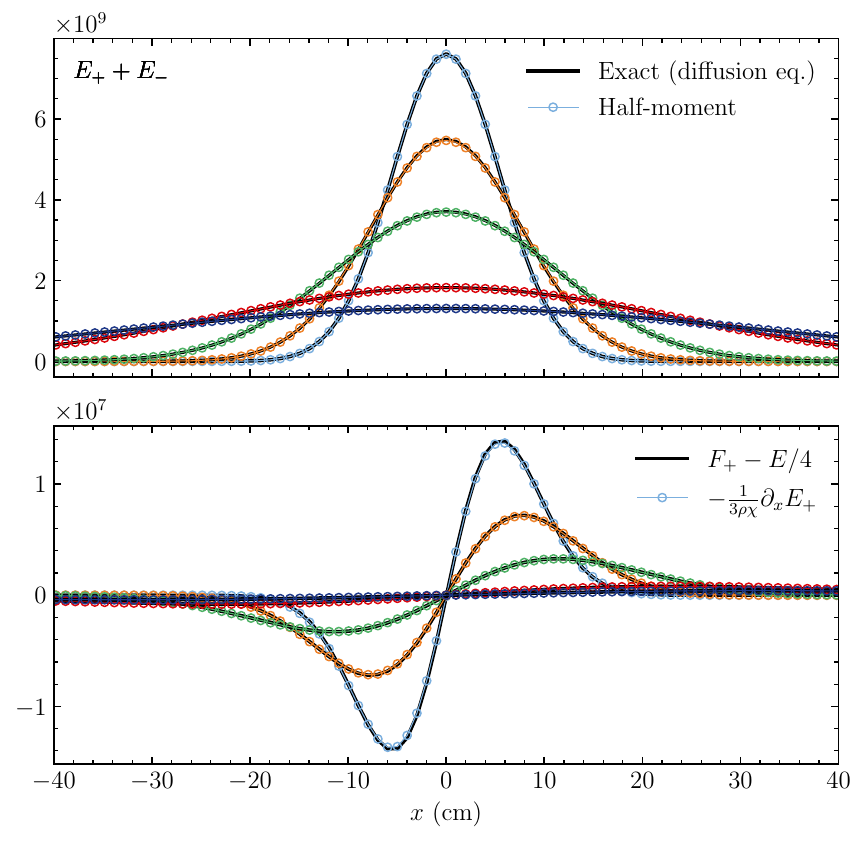}
\caption{Energy density (top) and diffuse flux (bottom) in the pure-scattering diffusion test. Thick black lines and thin lines correspond to the exact and HM solutions, respectively, shown at different times. Values are shown in erg cm$^{-3}$ at $t/(10^{-9}\,\textrm{s})=1$, $15$, $50$, $250$, and $500$.}
\label{fig:diff_scattering}
\end{figure}

In Figure \ref{fig:diff_scattering}, we compare the exact and HM solutions at different times, with $\hat{c}=c$. We obtain that both solutions match throughout the pulse's evolution, with typical errors of $0.1-1\%$ at early times and $0.01-0.1\%$ after $t\sim 250$ s. As is shown in the same figure for $\boldsymbol{F}_+$, Eq. \eqref{Eq:diffuseFpm} is satisfied at all times.

We also find that including the isotropy term ${\rho \sigma (E_\pm - E/2)}$ in the energy equation (Eq. \eqref{Eq:HMSourceTerms_grey}) is crucial for recovering the diffusion limit with pure scattering. Although Eq. \eqref{Eq:diffuseE} would also be obtained without that term, its omission produces a deviation between $E_+$ and $E_-$ while still keeping $\boldsymbol{F}_\pm$ close to $\pm (E/4)\hat{\boldsymbol{d}}$ according to Eq. \eqref{Eq:diffuseFpm}. This quickly leads to $f$ reaching values close to $1$, deviating from anisotropy and resulting in a departure of $\xi$ from its isotropic limit of $1/3$. Conversely, including the term guarantees that isotropy is maintained throughout the diffusion process, with $||\boldsymbol{F}_\pm||\approx E_\pm/2 \approx E/4$ and therefore $\xi=1/3$ at all times.

In this pure radiation case, where the cooling timescale depends on the free wave propagation speed with no influence of the gas internal energy (see Section \ref{SS:Diffusion_abs_scat}), the correct solution is only recovered when applying the speed limiter by \cite{Sadowski2013} (see Appendix \ref{A:SignalSpeeds}), which prevents excessive diffusion. This becomes more notorious as the optical depth is increased.

We also note small departures from the exact solution if a lower limit of $1/3$ is imposed on the Eddington factor $\xi$ as in \cite{Ripoll2005}. This can be understood by considering a slightly anisotropic radiation intensity of the form $I(\hat{\boldsymbol{n}})\propto (1 + a \hat{\boldsymbol{x}}\cdot\hat{\boldsymbol{n}})$ in the diffusion regime, leading to  $\xi = P^{xx}_{\pm}/E_\pm \approx (1/3\pm a/12)$ to first order in $a$. Thus, if $\xi\geq1/3$ is imposed, the $\pm a$ terms do not cancel out when summing the $+$ and $-$ HM equations, and Eq. \eqref{Eq:diffuseE} is not recovered. We conclude that such a floor value should not be applied in order to correctly recover the diffusion limit.

\subsubsection{Combined absorption and scattering}\label{SS:Diffusion_abs_scat}

We now consider a diffusion problem in a 1D medium able to absorb, emit, and scatter radiation. We impose LTE at $t=0$ with a temperature perturbed about the average value ${T_0=10^6}\,\mathrm{K}$ as ${T=T_0 \left(1 + A \sin(k x) \right)}$, where $A=10^{-3}$, $k=2\pi/L$, and $L=10$ cm. We consider the domain $[0,L]$ with a resolution of $101$ cells and periodic boundary conditions imposed on all fields. We define equal absorption and scattering opacities as $\kappa=\sigma=5$ cm$^2$ g$^{-1}$ and set a uniform density $\rho=1$ g cm$^{-3}$, in such a way that the total optical depth across the domain is $100$. 

A linearization of the internal energy and RTE equations for small $A$ \citep[see][Appendix B]{MelonFuksman2024partI} shows that, to first order, the initial perturbation decreases exponentially, with the radiative energy density evolving as
\begin{equation}
    E(t,x) = a_R T_0^4\left(
    1 + 4 A\, e^{-t/t_\mathrm{cool}}\sin(k x)
    \right)\,.
\end{equation}
In this expression,
\begin{equation}
    t_\mathrm{cool} = 
    \frac{\rho \epsilon}{4 a_R T_0^4}
    \frac{3 + \lambda_R \lambda_P k^2}{\lambda_R \lambda_P k^2}
    \frac{\lambda_P}{c}\,,
\end{equation}
where $\lambda_P=(\rho \kappa)^{-1}$ and $\lambda_R=(\rho \chi)^{-1}$ are the absorption and total mean free paths, respectively, while $\rho \epsilon = \frac{p}{\Gamma -1}$ is the gas internal energy, taking $\Gamma = 5/3$.

\begin{figure}[t!]
\centering
\includegraphics[width=\linewidth]{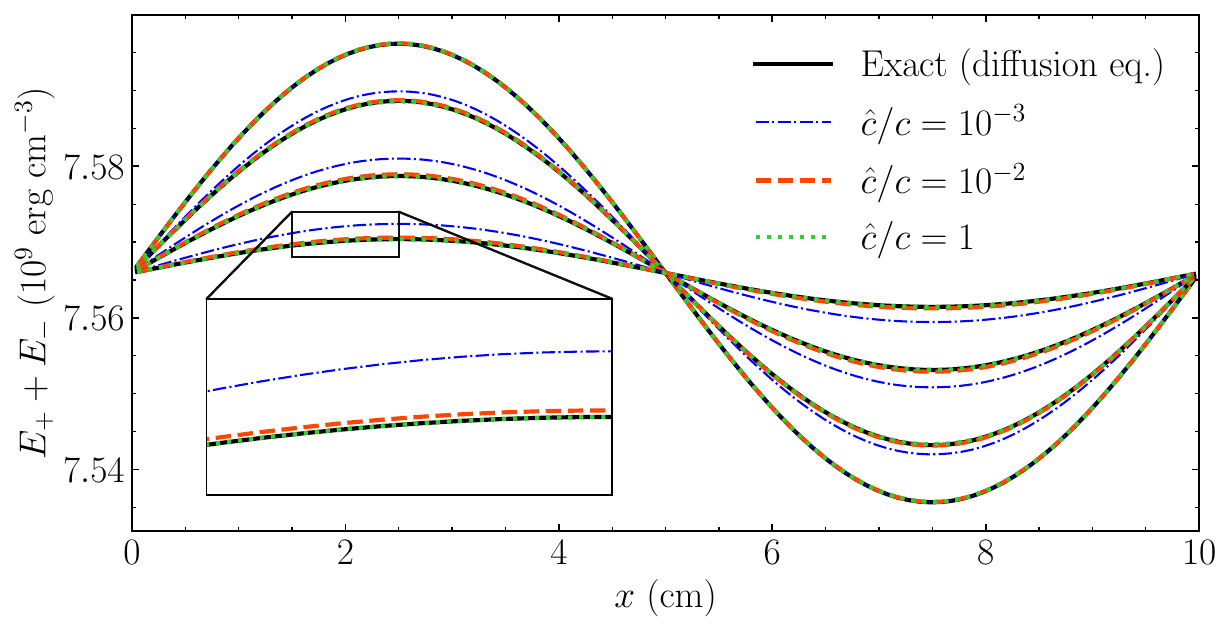}
\caption{Total radiation energy in the combined absorption and scattering diffusion test. Distributions are shown at $t/(10^{-6}\,\textrm{s})=0$, $3$, $9$, and $20$ for different values of  $\hat{c}/c$.}
\label{fig:diff_abs_scat}
\end{figure}

In this problem, the cooling timescale is unaltered when $\hat{c}<c$ for sufficiently high $\hat{c}$, as is shown in \cite{MelonFuksman2024partI}. This is because the radiation field perturbation can only decrease as the gas loses its internal energy perturbation with a cooling rate determined by $c(G^0_++G^0_-)$, which is independent of $\hat{c}$. If this cooling process is slow enough, the radiation fields become quasi-stationary, meaning that time derivatives can be dropped from 
Eq. \eqref{Eq:HMRT_multigroup}, resulting in a system independent of $\hat{c}$. In this case, the exact solution is recovered for $\hat{c}\gtrsim 10^{-2} c$, as is shown in Fig. \ref{fig:diff_abs_scat}. The property that diffusion times remain unchanged for high enough $\hat{c}$ is crucial for modeling systems with diffusion-regulated cooling, such as dense regions of circumstellar disks (Section \ref{S:Disks}).

\section{Disk models}\label{A:DiskModel}

\begin{figure}[t!]
\centering
\includegraphics[width=\linewidth]{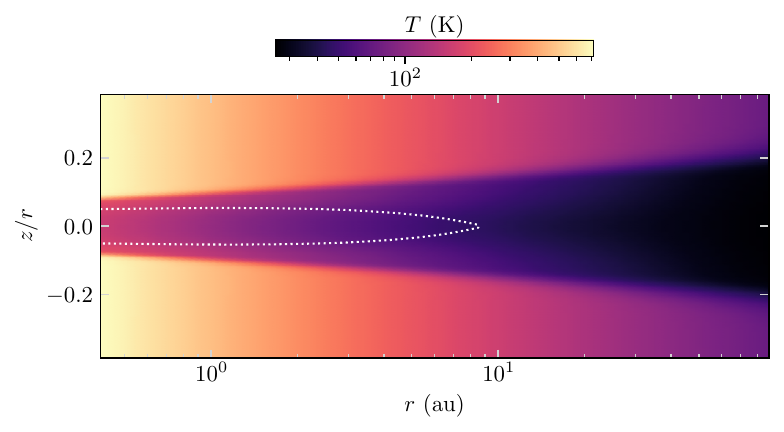}
\caption{
Equilibrium temperature in the HM protoplanetary disk model with $22$ frequency groups. The dotted line highlights the locations where the Planck vertical optical depth equals $1$.}
\label{fig:T_Disk}
\end{figure}

We define the gas density in our disk models as a distribution corresponding to vertically isothermal hydrostatic equilibrium, using the same stellar and disk parameters as in the initial distribution in Section 3 of \cite{MelonFuksman2022}.
As in that work, we neglect scattering and use the frequency-dependent dust absorption opacities obtained in \cite{Krieger2020,Krieger2022} for grains with sizes between
$5$ and $250$ nm, for which we assume a fixed dust-to-gas mass ratio of $f_\mathrm{dg}=10^{-3}$.
The opacities are tabulated for $132$ frequency values logarithmically sampled in the range  $[\nu_\mathrm{min},\nu_\mathrm{max}]=[1.5\times 10^{11},6\times 10^{15}]$ Hz. With these parameters, the maximum Planck vertical optical depth across the disk is $\tau_v\sim 300$, and the disk becomes optically thin in the vertical direction at $r\sim 8$ au (see Fig. \ref{fig:T_Disk}).

When computing the temperature with either the HM or M1 methods, we integrate the radiative transfer and gas energy evolution equations until thermal equilibrium is reached, keeping the mass distribution fixed. In these simulations, the gas internal energy evolves only due to radiation emission and absorption; that is, with $S^\mathcal{E}_\mathrm{HD/MHD}=0$ (Section \ref{S:HalfMoment}). A stellar irradiation source term is included as outlined in Section \ref{S:HalfMoment}, computed via ray tracing as in \cite{MelonFuksman2022} for all frequency bins in the opacity table. 

\begin{figure}[t!]
\centering
\includegraphics[width=\linewidth]{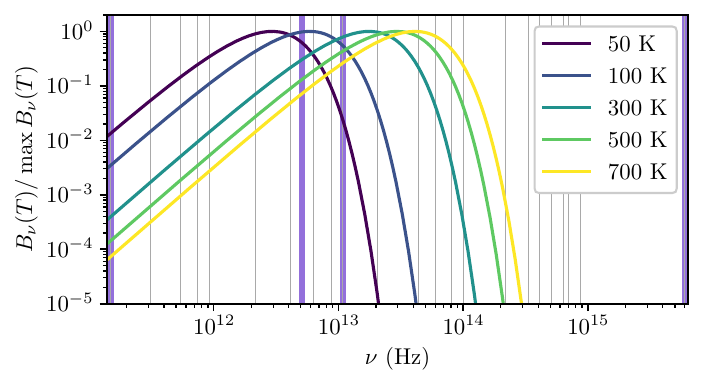}
\caption{Frequency groups used in the disk models. Thick purple lines delimit the $N_g=3$ groups, while thin gray lines delimit the groups for $N_g=22$. The Planck spectral radiance, normalized to have a maximum value of 1, is shown for different temperatures.}
\label{fig:groups}
\end{figure}

The radiation fields in the HM and M1 runs are defined using $N_g=1$, $3$, or $22$ groups, as is illustrated in Fig. \ref{fig:groups}. In our disk model, with maximum temperatures of $\sim 700$ K, the energy density in the highest-frequency groups for $N_g=22$ is several orders of magnitude lower than in the lower-frequency groups. We have chosen not to change this to test the robustness of the code under such conditions.

In all cases, equations are solved in spherical coordinates in the domain $(r,\theta)\in [0.4,100]\,\textrm{au}\times [\pi/2-0.4,\pi/2+0.4]$, using a grid resolution of $200\times 200$ and logarithmic spacing in $r$. For the HM runs, we define the splitting direction $\hat{\boldsymbol{d}}$ as the $\hat{\bm{\theta}}$ unit vector, which depends on the spatial location.
The disk is assumed to have an inner truncation radius of $r_c=10 R_s$, where $R_s=2.086 R_\odot$ is the assumed stellar radius. Thus defined, $r_c$ is smaller than the minimum radius of the domain, and so the optical depth of the disk portion left out of the domain must be added when computing the irradiation flux. For this computation, we follow the prescription in \cite{MelonFuksman2022}.

We fix zero-gradient conditions for the radiation fields at all boundaries except for $F^\theta_{+}$ and $F^\theta_{-}$ at the lower and upper $\theta$ boundaries, respectively, which are set in LTE at a temperature of $10$ K. To minimize energy accumulation at the upper $r$ boundary, we extrapolate the radiation energy densities at the corresponding ghost cells assuming a power law $\propto r^{-2}$. We use linear reconstruction everywhere except in the HM runs, where we switch to flat reconstruction if $f<0.47$. This minimizes spurious oscillations of the radiation field formed in this problem for $f<1/2$, guaranteeing that temperature variations stay below $0.1 \%$. 
Additionally, in the HM runs, we enforce Eq. \eqref{Eq:physlimit2} following the prescription in Eq. \eqref{Eq:Fminv2}. We justify this choice in Section \ref{SSS:DiskHM}.

Starting from a uniform temperature of $10$ K in LTE, the disk reaches thermal equilibrium once the stellar irradiation term balances diffusive cooling. With our disk parameters, it takes $\sim 100$ yr to reach a stable temperature distribution, with the exception of the outermost regions ($\gtrsim 50$ au) for $N_g=1$, which take up to $\sim 300$ yr. This delay is due to the opacity underestimation in those regions in the single-group case (Section \ref{SSS:DiskM1}), which increases the optically thin equilibration timescale. The same phenomenon can be seen in Fig. 8 of \cite{Pavlyuchenkov2025}. After equilibrium is reached, the temperature distribution is unchanged, and the values reported in Section \ref{S:Disks} are computed after $\sim 1000$ yr.

Unlike HM and M1, the MC method does not allow one to define an optical depth for disk portions excluded from the domain. Therefore, in the MC runs, we include the entire disk setting the inner boundary as the inner truncation radius. To do so, we define a new grid in the domain $(r,\theta)\in {[r_c,100\,\textrm{au}] \times [\pi/2-0.4,\pi/2+0.4]}$ with a resolution of $2500\times 200$ and logarithmic spacing in $r$. We set the total number of photon packages to $10^{10}$ and neglect scattering. We verified that small reductions in resolution and number of photons did not alter the obtained temperatures, nor did applying the Modified Random Walk method \citep{Fleck1984,Robitaille2010} for faster convergence in high-opacity regions.

For all methods, the final equilibrium temperature distribution differs from the one assumed to
define the
gas distribution. Nevertheless, this approach suffices for our goal of comparing temperature distributions obtained with different radiative transfer methods.

\end{appendix}

\end{document}